\documentclass[a4paper, 12pt]{article}
\usepackage{amsmath,amssymb,amsthm,amsfonts,array,mathrsfs,ifthen,mathtools}
\usepackage{microtype}
\usepackage{graphicx,epsfig}
\usepackage{lscape}
\usepackage{dcolumn}
\usepackage{eurosym}
\usepackage{enumitem}
\usepackage{multirow}
\usepackage{comment}
\usepackage[normalem]{ulem}

\usepackage[shadow,colorinlistoftodos]{todonotes}
\usepackage[toc,page,header]{appendix}
\usepackage{minitoc}
\usepackage{booktabs, makecell}
\usepackage[flushleft]{threeparttable} %notes in tables
\usepackage{lscape}
\usepackage{dcolumn}
\usepackage{chngcntr}

\usepackage[natbibapa]{apacite}
\DeclareGraphicsExtensions{.pdf,.png,.jpg}
\graphicspath{ {Images/} }
\usepackage{epstopdf}
\usepackage[singlespacing]{setspace}

\usepackage{relsize}

\usepackage{soul} % for underlined text

\definecolor{myblue}{RGB}{0,30,180}

\usepackage[colorlinks=true,linkcolor=black,
urlcolor=black,
citecolor=myblue,pdftex]{hyperref}

\usepackage[skip=0.333\baselineskip]{caption}
\usepackage{subcaption}
\newcolumntype{d}[1]{D{.}{.}{#1}}

\usepackage{float}
\usepackage{tikz}
\usetikzlibrary{arrows.meta, positioning,calc,decorations.pathreplacing}
\tikzstyle{process} = [
  rectangle, 
  rounded corners=4pt,
  minimum width=3.5cm, 
  minimum height=1.8cm, 
  text centered, 
  draw=black, 
  line width=0.8pt,
  top color=gray!10,
  bottom color=gray!20,
  text width=3.5cm, 
  align=center
]
\tikzstyle{arrow} = [very thick, ->, >=stealth]

\usepackage[sc]{mathpazo}
\usepackage{sectsty}
\usepackage{titlesec}

\titleformat{\section}
{\normalfont\Large\bfseries\centering\sc} % Format
{\thesection.}                          % Section number
{1em}                                   % Space between number and title
{}                                      % Code before the title

\titleformat{\subsection}
{\normalfont\large\bfseries\centering\sc}
{\thesubsection.}
{1em}
{}

\usepackage{authblk}

\begin{document}

\footnotesep=1.2\footnotesep
\baselineskip=1.2\baselineskip

\title{Beyond the Margin: Targeted Conservation and Household Water Demand \thanks{We thank Michele Garagnani, Tommaso Reggiani, attendants at the ESA European Meeting 2024 (Hanken School of Economics ), BEEN 2024 (University of Turin), Verona Experimental Meeting 2024 (University of Verona), and IMT Workshop on Behavioral and Experimental Economic 2024 (IMT School for Advanced Studies).\\
Andrea Albertazzi: IMT School for Advanced Studies Lucca, Lucca, Italy, andrea.albertazzi@imtlucca.it\\
Elisabetta Leni: Y-S\"{a}\"{a}ti\"{o}, Helsinki, Finland, elisabetta.leni@ysaatio.fi\\
Ennio Bilancini: IMT School for Advanced Studies Lucca, Lucca, Italy, ennio.bilancini@gmail.com} \\ 
}
\author[a]{Andrea Albertazzi}
\author[b]{Elisabetta Leni}
\author[a]{Ennio Bilancini}

\affil[a]{IMT School for Advanced Studies Lucca} %
\affil[b]{Y-S\"{a}\"{a}ti\"{o}}

\date{\today}

\maketitle

\begin{abstract}
\baselineskip=1.2\baselineskip
\noindent Non-price interventions targeting specific household water uses are increasingly central to conservation policy, but whether end-use savings translate into lower aggregate demand remains unclear. This paper reports evidence from a pre-registered field experiment in which 775 Finnish households were randomized to a shower timer, a water-saving shower head, or the same shower head with real-time feedback. Utility-grade water meters measure household-level effects, while shower-level data provide complementary end-use evidence for the two shower-head treatments. The shower timer has no detectable effect. In contrast, the water-saving shower head reduces daily household demand by about 5\%, and pairing it with real-time feedback doubles this reduction to about 10\%. The convergence between shower- and meter-based estimates shows that end-use savings are consistent with limited offsetting elsewhere in the home. Cost-benefit analysis indicates that combining technological constraint with salient point-of-use feedback dominates reminder-based strategies.
\\*[0.5\baselineskip]

\\*[0.5\baselineskip]
\end{abstract}

\newpage
\section{Introduction}

Freshwater scarcity is an increasingly important constraint on economic activity worldwide. Water stress is no longer confined to arid regions. Rapid urbanization, climate variability, and infrastructure deficits are making freshwater scarcity an increasingly urban and global phenomenon \citep{Biswas2025}. Reducing demand rather than expanding supply has accordingly become the organizing principle of water policy across many countries and regions. In the public water supply sector, households are the primary consumers. For instance, in the EU approximately 77\% of water supplied through public networks is consumed by households \citep{EEA2025}.\\
Price-based instruments have been the primary tool for regulating residential water consumption. However, a substantial empirical literature consistently documents that residential water demand is typically inelastic \citep{Dalhuisen2003, Worthington2008}. This has motivated a growing interest in non-price instruments, with evidence showing that norm-based messages and social comparison targeting overall household consumption can reduce residential water use \citep{FerraroPrice2013, FerraroMirandaPrice2011, bernedo2014persistent, BrentCookOlsen2015, Bhanot2017, Bhanot2021, JessoeetAl2021a}. However, the efficacy of such interventions has generally been limited, typically yielding modest reductions of approximately 3--5\%, and their effects tend to diminish over time.\\
A complementary strand of literature instead targets information provision directly at the point of use, specifically shower consumption. Within residential settings, showering alone accounts for between 25\% and 40\% of total indoor water use \citep{mazzoni2023, makki2013}, and because it is concurrently energy-intensive, habitual, and difficult to monitor in real time, it constitutes a policy lever with high environmental relevance and non-trivial economic implications. Consistent with this, recent studies document meaningful reductions in shower consumption of up to 22\% from targeted end-use interventions \citep[e.g.,][]{TiefenbeckEtAl2018,TiefenbeckEtAl2019}.\\
A central unresolved question is whether water savings induced by targeted end-use interventions translate into reductions in total household water demand. Savings at a single margin may have different implications for aggregate consumption if households adjust water use in other activities.\footnote{Such responses may attenuate the direct end-use effect if households increase water use in other water-intensive activities, such as laundry or cleaning, but may amplify it if the intervention increases attention to water use more broadly.} Whether end-use savings pass through to household demand is therefore both an empirical question and a policy-relevant issue.\\
This paper examines this pass-through in a pre-registered field experiment targeting one major residential end use: showering. A total of 775 households in Finland were randomized into a control group or one of three treatments: a shower timer, a water-saving shower head, or the same shower head paired with real-time consumption feedback. Leveraging utility-grade household water meters that separately record cold- and hot-water consumption, this paper estimates treatment effects at the household level. In parallel, using shower head-specific consumption measurements, it estimates the corresponding treatment effects at the point of use.\\
The results reveal a clear hierarchy across treatments. The shower timer produces no detectable effect on total household water consumption, suggesting that a low-cost behavioral nudge alone is insufficient to generate meaningful conservation at the household level. The water-saving shower head reduces daily household water use by approximately 9 liters (-5\%), and pairing it with real-time feedback roughly doubles this reduction to approximately 17 liters (-10\%). The shower-level data allow this pattern to be decomposed in two ways. First, restricting the comparison to households equipped with the same shower head, we find that feedback reduces water use per shower by approximately 7 liters (-13\%) relative to the no-feedback baseline, primarily through reductions in flow rate, with a modest additional contribution from shorter shower duration, and with no evidence of attenuation over time. This isolates the contribution of information provision from the mechanical effect of the device. Second, the implied household-level savings derived from the shower analysis closely match the meter-based estimates, with the latter capturing approximately 80\% of the former. This alignment between two data sources suggests that the savings generated at the point of use are consistent with limited offsetting elsewhere within the household.\\
A cost-benefit analysis that translates the estimated treatment effects into annual resource and bill savings shows that the shower head paired with real-time feedback is the only intervention that recovers its acquisition cost within the device's guaranteed service life. The water-saving shower head alone breaks even only over a 10-year horizon, while the shower timer is never cost-effective. These results imply that intervention design --- in particular, the pairing of efficiency technology with salient, real-time information --- matters at least as much as intervention cost for the welfare ranking of non-price conservation policies.

This paper contributes to three strands of literature. First, it speaks to a growing body of evidence on the aggregate effects of targeted behavioral interventions. A recurring finding in this literature is that interventions focused on a specific domain can trigger responses in non-targeted behaviors, with overall impacts that may differ substantially from estimates based on the targeted outcome alone \citep[e.g.,][]{Altmann2021, Medina2020}. In the context of residential water conservation, the evidence on cross-domain spillovers is mixed. \cite{Goetz2024} report that a hot water conservation intervention also reduced room heating energy consumption. \cite{JessoeetAl2021b} document an associated reduction in summertime electricity consumption following a social norm messaging campaign targeting residential water use. \cite{JAIMETORRES2018} find positive spillovers onto water consumption among untreated neighbors of households targeted by a social comparison intervention. \cite{BonanetAl2024} find that water conservation reports reduce both water and electricity consumption, but only among households not already receiving similar reports for other resources, suggesting that exposure to multiple nudges crowds out attention. \cite{Carlsson2021} document that a social information campaign targeting water use generates no average spillover on electricity, but produces meaningful reductions among households that were already efficient water users. Finally, \cite{LeongBuurmanTay2024} find that real-time feedback from smart shower devices reduces shower water use substantially, and indirectly estimate that approximately one third of these savings are eroded by compensatory increases in other water-related activities. Taken together, this evidence suggests that end-use conservation interventions need not be offset by compensatory responses elsewhere, and may in some cases generate positive spillovers onto related resource uses. However, the possibility of within-resource offset remains an open empirical question: savings at the targeted end use may be partly undone by increased consumption in other water-using activities. Our paper contributes to this literature by providing evidence consistent with limited offsetting of end-use savings in household water demand within a single experimental design.\\
Second, the paper contributes to the literature on the relative effectiveness of technological retrofits and behavioral information as non-price conservation instruments. \cite{TiefenbeckEtAl2018} show that real-time, activity-specific feedback on showering reduces household energy use more than aggregate electricity smart metering feedback administered to the same population, attributing both the magnitude and the immediate, stable onset of the effect to the correction of salience bias at the moment of resource use. Moreover, such an effect persists even in the absence of financial incentives \citep{TiefenbeckEtAl2019}. \cite{FangEtAl2023} document that real-time shower feedback and periodic shower energy reports jointly outperform either instrument in isolation, consistent with the two addressing distinct behavioral barriers: inattention and knowledge gaps about energy intensity, respectively. Whether real-time information similarly amplifies the effect of a mechanical flow restriction delivered through the same device has not been established. \cite{AnsinkOrnaghiTonin2021} study the technology and information components of a large-scale UK water audit program, finding that low-flow retrofit devices produce persistent household-level savings while the information component generates a larger initial response that subsequently fades. However, their design does not permit identification of the interaction between the two components. The current paper addresses this gap by randomly assigning households to use the same water-saving shower head with real-time feedback either on or off. This approach allows for a clear separation of the mechanical and informational effects of the device within a consistent, pre-registered experimental setup. Real-time feedback is found to intensify the mechanical effect, leading households to further reduce consumption beyond the baseline response, consistent with the salience mechanism.\\
Third, the paper contributes to the literature on the cost-effectiveness evaluation of non-price water conservation instruments. \cite{FerraroPrice2013} and \cite{TiefenbeckEtAl2018} provide cost-effectiveness calculations for individual non-price interventions, but neither compares alternative instruments within the same experimental design. This paper addresses that gap directly by translating the estimated treatment effects into annual water and energy bill savings, using the tariff structure faced by households in the sample, and computing net present values and payback periods for each of the three interventions relative to their retail acquisition costs. The resulting ranking is non-monotone in intervention cost. The shower timer, despite its negligible upfront expenditure, fails to recover even that cost because it generates no measurable savings. The water-saving shower head alone reaches break-even only over a horizon that exceeds its guaranteed service life at the retail price of the smart device used in this study. In contrast, the shower head paired with real-time feedback generates a positive net present value within five years and is the only intervention that is privately cost-effective on water bill savings alone.\\
More broadly, the paper emphasizes that the policy relevance of targeted conservation interventions depends not only on their ability to modify behavior at the treated margin but also on the extent to which these behavioral adjustments persist when evaluated in terms of overall household demand. In this regard, the empirical evidence yields a practical implication for conservation policy: interventions that jointly employ technological constraints and salient, point-of-use informational cues may constitute a more effective pathway to achieving substantive reductions in household consumption than strategies relying solely on low-cost, reminder-based approaches.\\

The remainder of the paper is organized as follows: Section \ref{sec:design} introduces the institutional setting and experimental design; Section \ref{sec:results} presents the main findings; Section \ref{sec:policy} provides a cost-benefit analysis of our findings; and Section \ref{sec:conclusions} concludes. An additional appendix supplements this document.
\vspace{0.5cm}

%%%%%%%%%%%%%%%%%%%%%%%%%%%%%%%%%%%%%%%%%%%%%%%

%%%%%%%%%%%%%%%%%%%%%%%%%%%%%%%%%%%%%%%%%%%%%%%%%%%

\section{Experiment}\label{sec:design}
This section summarizes the institutional setting, the water-saving devices used, and further experimental details. The study was pre-registered at the OSF Registry\footnote{The pre-registration is available at the following link: \href{https://doi.org/10.17605/OSF.IO/BJVZD}{https://doi.org/10.17605/OSF.IO/BJVZD}.}, and conducted as part of Y-S\"{a}\"{a}ti\"{o} activities to promote sustainability.\footnote{For more information on this matter please visit \href{https://ysaatio.fi/en/about-us/sustainability/}{https://ysaatio.fi/en/about-us/sustainability/}.}\\

\subsection{Institutional Setting}
The intervention was implemented in a sample of 16 government-subsidized properties (all apartment buildings) owned by \href{https://ysaatio.fi/en/}{Y-S\"{a}\"{a}ti\"{o}}, Finland’s largest national non-profit landlord and a major provider of affordable rental housing. Rents in these apartments are set below prevailing market rates (approximately 30\% lower). The selection of tenants follows criteria and rules set by the state that prioritize people in urgent need of housing and low-income earners. In such state-subsidized housing, a third of the households are in the lowest income quintile, and three out of four are below the median. Around 5\% are in the top two income quintiles. Not significantly different from those living in rental market.\footnote{Information retrieved from \cite{Toikka2025}.} Consequently, the experimental participants are not a representative sample of the Finnish population but mostly represent the bottom half of the income distribution in the country. Average monthly tenant turnover rate is approximately 6\%.\\
The sampled properties exhibit substantial heterogeneity in scale (18-119 apartments), construction year (1950-2022), location, and metering infrastructure. Approximately 85\% of households have apartment-level water meters that measure individual water use. For these households, water bills are calculated separately for cold and hot water, thereby providing the correct price signal. The remaining households rely on a single property-level meter, which records aggregate water use only and does not permit observation of apartment-specific consumption. In these properties, water charges are allocated on a per-unit basis by dividing the property’s total consumption costs equally across registered tenants. Across all properties, hot water and space heating are provided centrally. The heat supply is sourced from district heating that utilizes waste heat recovered from cooling water discharged by power plants.

\subsection{Devices and randomization}
\begin{figure}[h]
	\centering
\includegraphics[width=.5\linewidth]{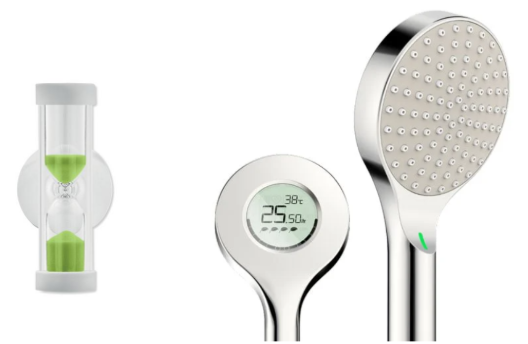}
	\caption{Shower timer (left) and smart shower (right).}\label{fig:devices}
\end{figure}

This study explores whether interventions targeting a single major end use (showering) ultimately lead to sustained savings at the household level. To this purpose, we provided households with two typical water-conservation devices designed to reduce shower water consumption: a shower timer and a water-saving hand shower. Both devices are depicted in Figure \ref{fig:devices}.\\

The shower timer used in this study is a simple, inexpensive device that helps track shower time. The device features a suction pad that allows it to be easily hung in the shower, and it empties in approximately 4 minutes.\\
The smart hand showers were supplied by \href{https://www.oras.com/en/home}{Oras}, a Finnish provider of sanitary fittings. The device is a digital hand shower equipped with an integrated display powered by the water flow. During use, it provides real-time feedback on cumulative water consumption, associated energy use, and water temperature, and includes a colored LED indicator intended to signal the sustainability of the ongoing shower. The hand shower is Bluetooth-enabled, enabling remote connectivity. Moreover, because these devices incorporate a flow controller, their flow rate was set to approximately 9 liters per minute, which may mechanically reduce water use relative to the fixtures they replaced, which had a flow rate of about 12 liters per minute.\\
To identify whether interventions targeted at a single major end use, i.e., showering, translate into lower total household water demand, we randomly assigned properties (and therefore households within them) to one of four experimental arms. In \textbf{Control (\textit{C})}, tenants received no device. This arm serves as the baseline against which we estimate the effects of the shower timer and digital shower interventions. In treatment \textbf{Timer (\textit{T})}, tenants received a shower timer by mail along with written instructions for use. This arm captures a low-cost reminder-based intervention that may affect shower duration, but requires attention and active compliance. In the first shower treatment, \textbf{Shower with No Feedback (\textit{SNoF})}, a digital hand shower was installed in the apartment, but the display only showed water temperature, and the sustainability indicator was disabled. As a result, tenants in this group received a water-saving device without any real-time feedback. This arm isolates the effect of the technological retrofit (flow restriction) without real-time behavioral feedback. Finally, in \textbf{Shower with Feedback (\textit{SF})}, a digital hand shower was installed in the apartment, and during the first ten showers, the device displayed only water temperature, mirroring the \textit{SNoF} condition. This phase provides a common baseline for showering behavior among apartments equipped with the digital shower. Beginning with the eleventh shower, tenants in this arm started receiving real-time feedback on shower behavior, including water and energy use, temperature, and an overall sustainability indicator. This arm, therefore, captures the effect of adding point-of-use feedback to the same underlying water-saving device.\\
To ensure that randomization produced groups with similar characteristics, the selected properties were stratified by dimension and by whether apartments had a household-specific water meter installed. Properties were randomly assigned to the four experimental groups within these strata. Table \ref{tab:buildings} summarizes the randomization across property type and treatments.

\begin{table}[h]\small\centering\small
\def\sym#1{\ifmmode^{#1}\else\(^{#1}\)\fi}
\begin{tabular}{@{}cccccc@{}}
& & \multicolumn{4}{c}{Intervention}\\
\cmidrule(l){3-6} 
\begin{tabular}[c]{@{}c@{}}Water meter\\installed\end{tabular} & Property & \textit{C} & \textit{T} & \textit{SNoF} & \textit{SF}\\
\midrule
NO & 4 small & 1 & 1  & 1  & 1 \\ 
\cmidrule(l){3-6} 
YES & 4 small   & 1  & 1  & 1   & 1 \\ 
\cmidrule(l){3-6} 
\multirow{2}{*}{YES}     & \multirow{2}{*}{8 medium-large} & 2 & 2 & 1 & 1 \\
 &  & 0  & 0 & 1\sym{\star} & 1\sym{\star} \\
\bottomrule
\end{tabular}
\caption{Properties characteristics across interventions. This table summarizes the treatment assignment across properties, along with their size, and whether apartments have a water meter installed. \sym{\star} in these properties, randomization occurred at the floor level. A property is classified as small if it contains no more than 45 apartments. Two small properties assigned to $C$ and $T$ were included in the broader Y-S\"{a}\"{a}ti\"{o} sustainability rollout, but not in the estimation sample. Because neither meter nor shower-level data are available for these properties, they are excluded from the current analysis.}\label{tab:buildings}
\end{table}

As reported in Table \ref{tab:buildings}, one property assigned to \textit{SNoF} and one assigned to \textit{SF} were instead randomized at the floor level. Within each of these two properties, we randomly allocated half of the apartments on each floor to \textit{SF} and the remainder to \textit{SNoF}. This within-building randomization holds constant all time-invariant property characteristics, allowing us to assess whether the effect of feedback on shower consumption persists after netting out these characteristics, and to test for spillovers between treated and untreated units within the same building.

\subsection{Additional Details}
All households, regardless of experimental assignment, received generic information about the ongoing study via letter.\footnote{See Appendix \ref{app:other_details} for the letters delivered to households in each treatment.} Households were defaulted to participate and could withdraw from the study via explicit request. Of the 775 targeted households, 109 (14.06\%) did not have an installed water meter and are therefore excluded from the household-level analysis. Among the remaining 666 households, 79 (11.86\%) were vacant for all or part of the study period; 11 (1.65\%) have been excluded because they either denied access to the apartment, opted out, had the shower head removed or broken, or reported a faulty water meter; and 3 (0.45\%) have no point observations in the pre- and/or post-intervention period.\\
Prior to the intervention, the digital-shower supplier installed communication modems at the properties assigned to the \textit{SNoF} and \textit{SF} arms. These modems collect usage data transmitted from the shower units via Bluetooth and relay the information to a cloud-based platform, enabling the research team to monitor outcomes in near real time. To mitigate the risk of tenant interference (e.g., unplugging), the modems were placed in common areas such as corridors and were typically concealed from view or installed in less accessible locations. Installations of the digital shower devices took place between September 25 and October 4, 2023. Over the same period, tenants in the \textit{T} arm received the shower timer by mail. Y-S\"{a}\"{a}ti\"{o} staff delivered the letters and shower timers, while the technical team left a copy of the letter in the apartments at the time of shower installation. Please refer to Table \ref{tab:delivery_dates} in Appendix \ref{app:other_details} for a breakdown of delivery dates for the information letters for each property, i.e., start of the intervention.\\
Before the start of the intervention and after the end of the study, we administered two surveys to investigate attitudes, behaviors, and beliefs that may influence water consumption and help us interpret the results. Because the number of households responding to both survey waves is small, the survey data are not sufficiently informative for the main analysis. Selected summary statistics are therefore reported in Appendix \ref{app:add_results}.\\
Figure \ref{fig:timeline} summarizes the treatment arms, the intervention duration considered in our analysis, the data used for treatment comparisons of household water consumption, and the data used for the shower-level analysis.

\begin{figure}[h]
  \centering 
  \resizebox{0.8\textwidth}{!}{%
\begin{tikzpicture}[
  font=\sffamily,
  box/.style={
    draw=gray!60,
    fill=gray!60,
    rounded corners=12pt,
    minimum width=6.2cm,
    minimum height=1.25cm,
    align=center,
    text=white,
    inner sep=6pt
  },
  bigbox/.style={
    draw=black,
    fill=gray!60,
    rounded corners=18pt,
    minimum width=3.8cm,
    minimum height=6.0cm,
    align=center,
    text=white,
    inner sep=8pt
  },
  highlight/.style={
    box,
    draw=black,
  },
  braceR/.style={
    decorate,
    decoration={brace,amplitude=6pt},
    thick
  }
]

% Left block (Before)
\node[bigbox] (props) {16 properties\\(N=775)};

% Right blocks (After)
\node[highlight, right=1.3cm of props.north east, anchor=north west] (control) {\textit{C}\\(N=236)};
\node[highlight, below=3mm of control] (hourglass) {\textit{T}\\(N=176)};
\node[highlight, below=3mm of hourglass] (showerNF) {\textit{SNoF}\\(N=179)};
\node[highlight, below=3mm of showerNF] (showerF) {\textit{SF}\\(N=184)};

% Braces + labels
\draw[braceR] ($(showerNF.north east)+(0.45,0)$) -- ($(showerF.south east)+(0.45,0)$)
  node[midway,right=10pt,align=center,font=\small,text=black] {Shower\\data};

\draw[braceR] ($(control.north east)+(2.3,0)$) -- ($(showerF.south east)+(2.3,0)$)
  node[midway,right=10pt,align=center,font=\small,text=black] {Meters data\\(if installed)};

% Timeline arrow + labels
\coordinate (baseL) at ($(props.south west)+(0,-1.0cm)$);
\coordinate (baseR) at ($(showerF.south east|-baseL)$);
\draw[->,thick] (baseL) -- (baseR);

% --- tick at start of intervention (between Before and After) ---
\coordinate (t0x) at ($(props.south east)!0.5!(control.south west)$); % midpoint of the gap
\coordinate (t0)  at (t0x|-baseL);                                   % project down to the arrow
\draw[thick] ($(t0)+(0,3pt)$) -- ($(t0)+(0,-3pt)$);
% optional label:
\node[below,font=\small,align=center] at ($(t0|-baseL)$) {Start of\\intervention};

\node[below,font=\small,align=center] at ($(props.south|-baseL)$)
  {Before\\(1 month)};

\node[below,font=\small,align=center] at ($(control.south|-baseL)$)
  {After\\(3 months)};

\end{tikzpicture}
}
  \caption{Summary of the study design. This figure shows the actual number of households (N) assigned to each treatment condition, the intervention timeline, and the corresponding data used in the analysis.}\label{fig:timeline}
\end{figure}
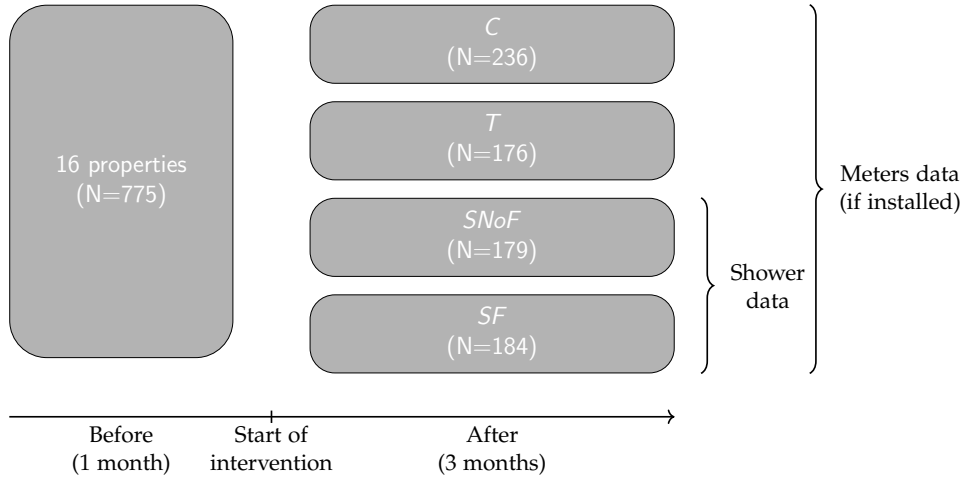

\subsection{Estimation and Inference}
Following our preregistration, we report results based on the prespecified observation window, spanning one month before the intervention and three months after.\\
First, using daily water meter readings, we estimate treatment effects on household water consumption for apartments assigned to \textit{T}, \textit{SNoF}, and \textit{SF}, relative to the control arm (\textit{C}). Consistent with the preregistered plan, we construct a balanced household panel by retaining for each household the 20 \textit{useful} daily readings immediately preceding the intervention start date and the 60 \textit{useful} daily readings immediately following it. This approach restricts the sample to households with at least this many observations in each phase. We define a daily read as \textit{useful} if cold-water consumption is at least 10 liters, a threshold intended to exclude days when the apartment is likely unoccupied. We find no evidence that treatment assignment affects the number of \textit{useful} readings ($F$-test: $p=0.440$).\\
Second, we present the shower-level analysis to integrate our main result on household water consumption. As preregistered, we restrict the sample to apartments with a smart hand shower installed (\textit{SNoF} and \textit{SF}) and for which we observe the full baseline of 10 showers and 60 subsequent showers. Unanticipated connectivity issues (i.e., range limitations) prevented approximately one-third of shower records from being uploaded to the cloud, resulting in missing timestamps for these events. The corresponding non-timestamp data were recovered manually by operators. This has two implications. First, in the shower-level analysis, we use the sequential shower index as a proxy for time. Second, we extend the observation window through the dates on which the data were retrieved. Doing so ensures that showers without timestamps occurred prior to retrieval. Extending the observation window to the retrieval dates is unlikely to affect our preregistration plan, since, with an average of 1.4 showers per household per day, the 70 shower events used in the analysis are expected to fall within the specified window (three months post-intervention).\footnote{The average number of daily showers is computed from shower events with timestamps. Because connectivity issues are orthogonal to households’ behavior and characteristics, we expect no systematic differences in showering behavior between observations with timestamps and those without.}\\
In Appendix \ref{app:add_results}, we replicate the analysis using the full set of available observations for both households and shower consumption data. The resulting estimates are comparable to those in the main text and leave our qualitative conclusions unchanged, indicating that the findings are not driven by the pre-registered sample restrictions.\\
The main specifications report standard errors clustered at the household level to accommodate correlation across repeated observations within apartments. Because most treatment assignments occurred at the property level, household-level clustering does not capture arbitrary correlation in consumption changes across apartments within the same property. At the same time, the small number of properties limits the reliability of conventional property-clustered standard errors. We therefore complement the regression estimates with a property-level diagnostic (Figure \ref{fig:property_water_change}). We plot the mean within-household change in daily water consumption for every property-treatment cell. The figure aligns with the findings reported in the results section. This indicates that the estimated magnitudes are not driven by a single property or by differences in property size. Nevertheless, this diagnostic addresses the influence of individual properties rather than fully resolving the inferential limitation posed by the small number of property-level assignments. The household-clustered $p$-values should therefore be interpreted with this qualification in mind. We are also unaware of any major property-specific shocks during the study period.
%%%%%%%%%%%%%%%%%%%%%%%%%%%%%%%%%%%%%%%%%%%%%%%%%%%%%%

\section{Results}\label{sec:results}

\subsection{Household Demand} \label{sec:meters}
This section examines whether intent-to-treat interventions aimed at decreasing shower water use significantly impact overall household water consumption. Indeed, a reduction in one end-use need not reduce aggregate demand if households reallocate water toward other activities. In particular, tenants may treat water saved in the shower as slack that can be redeployed to other water-intensive tasks, thereby changing the composition of water use rather than its level. This possibility is especially relevant in our setting, where the feedback targets a single, salient end use but does not mechanically constrain or price other uses. Using household-level meter readings, we test whether the three technologies, a shower timer ($T$), a water-saving shower head without feedback ($SNoF$), and the same shower head paired with real-time feedback ($SF$), translate into changes in overall household water use.

\begin{table}[h]\centering\footnotesize
	\def\sym#1{\ifmmode^{#1}\else\(^{#1}\)\fi}
	\begin{tabular}{l*{1}{cccccc}}
\hline
\hline
 & $C$ & $T$ & $SNoF$ & $SF$ & $p$-value \\
\midrule
Water (L) & 155.08 (109.25) & 193.42 (127.08) & 180.38 (114.87) & 191.00 (122.18) &  0.196 \\
\midrule
Number of tenants &   1.33 (  0.66) &   1.61 (  0.84) &   1.46 (  0.89) &   1.70 (  0.91) &  0.119 \\
Area (m$^2$) &  44.67 ( 11.01) &  47.41 ( 13.95) &  44.34 ( 15.30) &  47.35 ( 14.66) &  0.634 \\
Floor &   3.87 (  2.11) &   4.16 (  1.82) &   3.74 (  1.78) &   2.59 (  1.27) &  0.000 \\
\midrule
Households & 172 & 116 & 119 & 126 & 533 \\
\hline\hline
\end{tabular}\caption{Average households' daily water consumption in the baseline phase, along with household characteristics, by treatment. Standard deviations in parentheses. The last column shows $p$-values from joint \textit{F}-tests of the hypothesis that the interventions are not correlated with the control group.}\label{tab:meter_baseline}
\end{table}

\begin{figure}[H]
	\centering
\includegraphics[width=0.8\linewidth]{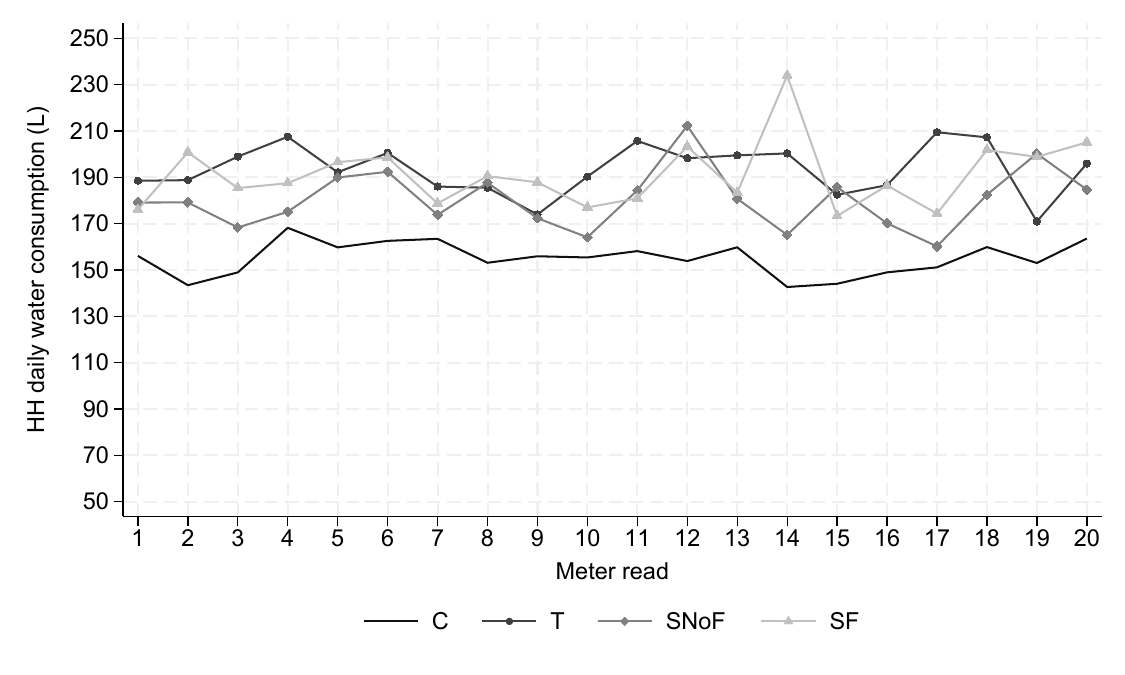}
\caption{Households' daily water consumption (L) before intervention by treatment arm.}\label{fig:meters_baseline}
\end{figure}

Table \ref{tab:meter_baseline} reports average daily water consumption and baseline household characteristics by experimental group during the pre-intervention phase. Baseline water use is not statistically different across groups ($p=0.196$), and observable characteristics are broadly balanced, consistent with successful random assignment. While floor level differs across groups, it is time-invariant and absorbed by household fixed effects. Our identification relies on differential changes in consumption between the intervention and baseline periods in treated households relative to the control group, while further controlling for meter-read fixed effects.\\
In addition to average water consumption, we assess whether the four treatment arms exhibit similar trends during the baseline period prior to intervention. Figure 
\ref{fig:meters_baseline} shows that consumption trends are comparable across conditions, and we find no evidence of linear differences in pre-trends between arms (\textit{F}-test, $p = 0.525$).\footnote{Please see Appendix \ref{app:add_results} for the estimation procedure.}

\begin{figure}[h]
	\centering
    \includegraphics[width=0.8\linewidth]{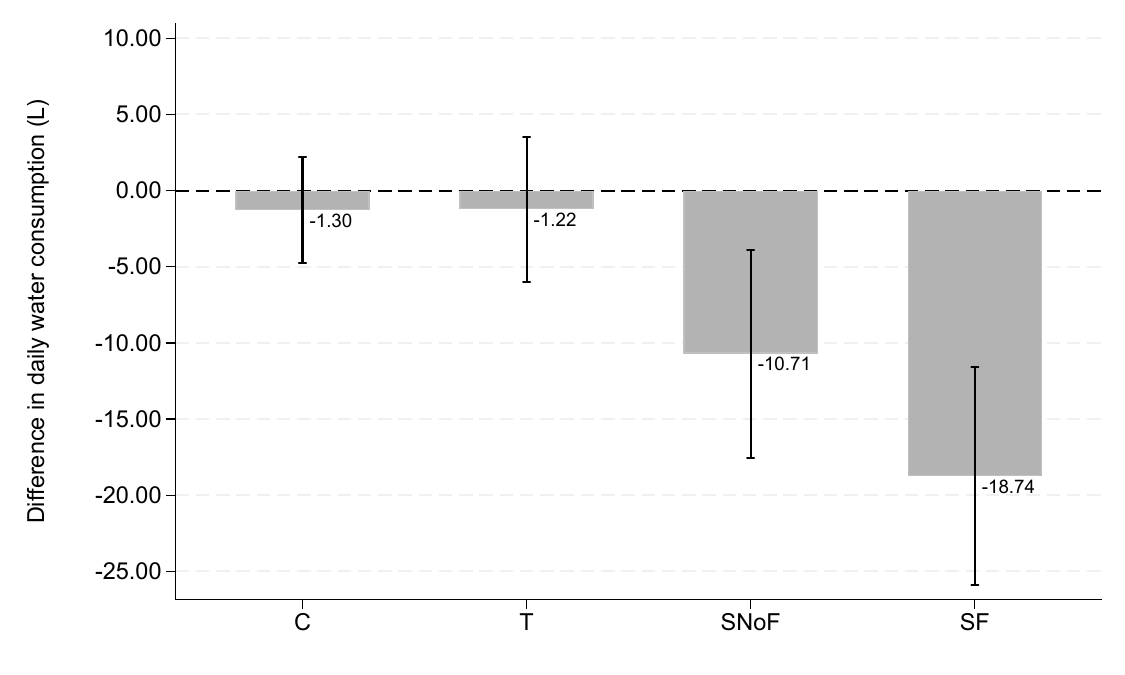}
	\caption{Mean within-household change in daily water consumption by treatment. Vertical solid lines represent 95\% confidence intervals.}
    \label{fig:diff_in_total}
\end{figure}

Figure \ref{fig:diff_in_total} provides a nonparametric summary of within-household changes in daily water consumption between the intervention and the baseline phase. Total consumption is essentially unchanged for households in the control group ($C$) and for those assigned to the timer treatment ($T$), suggesting that the device does not generate meaningful water conservation. Because the shower timer installation and use are not observed, the comparison identifies the causal effect of assignment to the timer condition (an intention-to-treat estimate), rather than the effect of actual use of the timer. In contrast, households with a water-saving shower head installed exhibit significant reductions in daily water consumption. The installation of the shower head alone ($SNoF$) is associated with a reduction of 10.71 liters per day. When the same device is paired with real-time feedback ($SF$), the reduction increases to 18.74 liters per day.\\
We formalize these comparisons by estimating the following intent-to-treat specification:
\begin{equation}\label{eq:meters}
    w_{it} = \alpha_{i} + \beta_{1} T_{it} + \beta_{2} SNoF_{it} + \beta_{3} SF_{it} + m_{t} + \epsilon_{it},
\end{equation}
where $w_{it}$ is household $i$'s daily water consumption in meter read $t$. The indicators $T_{it}$, $SNoF_{it}$, and $SF_{it}$ equal zero for all meter reads $t \leq 20$ and switch to one thereafter for households assigned to the corresponding treatment arm. The specification includes household ($\alpha_i$) and meter-read fixed effects ($m_t$) to absorb time-invariant heterogeneity and common shocks.\footnote{As preregistered, we index time by the meter-read sequence and include meter-read-number fixed effects, rather than calendar-date fixed effects. Because \textit{useful} meter reads are staggered across households, calendar-date fixed effects might create identification problems, leading to the loss of some observations. Using meter-read-number fixed effects captures common time variation while retaining these observations. This choice is unlikely to matter quantitatively, as the meter-read number is highly correlated with calendar time (Pearson's correlation coefficient: $r = 0.9877$). In Appendix \ref{app:add_results}, we perform the same analysis using all data and calendar-date fixed effects, and find that our main results still hold.} Therefore, coefficients $\beta_{1}$, $\beta_{2}$, and $\beta_{3}$ can be interpreted as the difference in consumption between the \textit{T}, \textit{SNoF}, and \textit{SF} treatment respectively, compared to the control group.

\begin{table}[h]\centering\small
\def\sym#1{\ifmmode^{#1}\else\(^{#1}\)\fi}
\begin{tabular}{l*{3}{cc}}
\hline\hline
          &\multicolumn{1}{c}{Total}&\multicolumn{1}{c}{Cold}&\multicolumn{1}{c}{Hot}\\
\hline
\textit{T} &0.0806 (2.98)         &-1.903 (1.84)         &1.984 (1.51)         \\
\textit{SNoF} &-9.405\sym{*} (3.86)         &-5.103\sym{*} (2.29)         &-4.302\sym{*} (1.90)         \\
\textit{SF} &-17.44\sym{***} (4.01)         &-8.741\sym{***} (2.28)         &-8.695\sym{***} (2.01)         \\
Costant    &176.6\sym{***} (1.30)         &111.5\sym{***} (0.79)         &65.14\sym{***} (0.63)         \\
\hline
$\hat\beta_{1} = \hat\beta_{2}$&   0.024         &    0.192         &  0.003         \\
$\hat\beta_{1} = \hat\beta_{3}$&0.000         &  0.005         &0.000         \\
$\hat\beta_{2} = \hat\beta_{3}$&    0.107         &    0.193         &   0.081         \\
$\hat\beta_{3} < \hat\beta_{2}$&        0.054     &   0.097    &   0.041   \\
\midrule
Households&      533         &      533         &      533         \\
Observations&    42,640         &    42,640         &    42,640         \\
\hline\hline
\end{tabular}
  \caption{Effect of interventions on households' daily water consumption. The first column reports estimated coefficients for total daily water consumption. The last two columns report results from the estimated model for daily cold and hot water consumption only. Standard errors clustered at the household level in parentheses. \sym{*} \(p<0.05\), \sym{**} \(p<0.01\), \sym{***} \(p<0.001\).}
  \label{tab:meter_read_all_pre_reg}
\end{table}

Table \ref{tab:meter_read_all_pre_reg} reports the estimated treatment effects of the interventions on households' daily water consumption. The estimates indicate that the shower timer treatment (\textit{T}) has essentially no effect on total daily water use: the point estimate is small and positive, and economically negligible. By contrast, both treatments involving the water-saving shower head reduce household demand substantially. Installing the shower head without feedback (\textit{SNoF}) lowers daily water consumption by 9.4 liters, whereas pairing it with real-time feedback (\textit{SF}) yields a reduction of 17.4 liters per day. Relative to the estimated baseline level in the control group, these magnitudes correspond to declines of approximately 5 percent under \textit{SNoF} and 10 percent under \textit{SF}. The equality tests at the bottom of the table reinforce this pattern. Both shower-head treatments differ from the timer arm ($p=0.024$ for $\hat\beta_1=\hat\beta_2$ and $p<0.001$ for $\hat\beta_1=\hat\beta_3$), indicating that the reduction in household demand is driven by the installation of the water-saving device rather than by the low-cost reminder alone. At the same time, while the incremental effect of adding feedback to the water-saving shower head is economically large, it is not statistically distinguishable from zero at conventional levels under a two-sided test ($p=0.107$ for $\hat\beta_2=\hat\beta_3$). The larger variance in these two treatments likely reflects heterogeneity in shower head usage across days, as household-level consumption depends on whether and how the shower head is used on a given day. The point estimates nevertheless suggest an economically meaningful incremental effect of real-time feedback. Because the comparison between \textit{SF} and \textit{SNoF} isolates the added effect of feedback on the same underlying device, and prior evidence points to a directional effect whereby feedback should significantly reduce consumption (as also confirmed by our estimates in Section \ref{sec:showers}), we further report the corresponding one-sided test of $H_0:\beta_3 \geq \beta_2$ against $H_1:\beta_3 < \beta_2$. Under this directional test, the incremental effect of real-time feedback is weakly significant ($p=0.054$).\\
The decomposition of total water use into hot and cold components provides additional insight into the nature of these aggregate effects. Both shower-head treatments reduce hot-water consumption, implying that the intervention affects the energy-intensive component of household demand rather than only total volumetric use. The estimated reduction in daily hot-water use is 4.3 liters under \textit{SNoF} and 8.7 liters under \textit{SF}, hence the point estimate for the feedback treatment is roughly twice as large. Although the difference between these two coefficients is only marginally significant under a two-sided test ($p=0.081$), the magnitudes are consistent with feedback strengthening the effect of the water-saving device on the energy-relevant margin. Given the directional hypothesis that real-time feedback should reduce consumption relative to the same device without feedback, the corresponding one-sided test yields significance at the 5 percent level ($p=0.041$). This is economically important because lower hot-water use not only saves water but also reduces energy demand.\\
Cold-water consumption also declines under both shower-head treatments. This pattern helps rule out a purely compositional interpretation of the results. In particular, if households were merely adjusting the temperature mix of water use while keeping total volume broadly unchanged, one would expect reductions in hot water to be offset by increases in cold water. Instead, both components fall. The evidence therefore points to a genuine reduction in total household water demand, affecting both its volumetric and its energy-intensive components. Overall, the findings suggest that the water-efficient shower head achieves substantial reductions in household resource consumption and that integrating real-time feedback into the device can further amplify these conservation effects. On the other hand, the shower timer intervention does not produce noticeable changes in total water demand.\\
In Appendix 
\ref{app:add_results}, we conduct the same analysis with two robustness checks. First, using all observations available for households that do not move within the four-month pre-registered period, and second, using all observations available up to six months after the intervention date. As new tenants were not informed about the ongoing study and were not given the opportunity to withdraw, we use only observations from tenants who were originally included in the study. These robustness checks indicate that the intervention effects observed in the main analysis are not reliant on the pre-registration plan and remain consistent over an extended period of up to 6 months. 

\subsection{From Shower Behavior to Household Demand}\label{sec:showers}
Using data on shower consumption, which in our sample accounts for 34\% of the baseline residential water use, this section supports our main result by isolating the incremental effect of real-time feedback while holding the shower device constant. Table \ref{tab:shower_baseline} presents summary statistics for the baseline phase, when none of the households receive real-time feedback. The two central columns provide mean values and standard deviation in parentheses for each variable. The last column provides the \textit{p}-values from \textit{t}-tests for treatment differences. The statistics show that before the onset of real-time feedback, shower behavior is comparable across arms in all but one variable. The flow rate is significantly higher in \textit{SF}, but this is likely because households in this treatment are, on average, one floor lower than those in \textit{SNoF} and therefore have higher water pressure. No other meaningful differences across treatment arms are detected.
%We have about 65\% of showers with a timestamp. Average number of showers per day: 1.401677 (0.7992918) for all apartments; and 1.317943 (0.7190938) for apartments with only one tenant.

\begin{table}[h]\centering\small
	\def\sym#1{\ifmmode^{#1}\else\(^{#1}\)\fi}
	\begin{tabular}{l*{1}{ccc}}
		\hline\hline
		& \multicolumn{2}{c}{Shower treatment} &\multicolumn{1}{c}{}	\\
        \cline{2-3}
		&\multicolumn{1}{c}{\textit{SNoF}}&\multicolumn{1}{c}{\textit{SF}}&\multicolumn{1}{c}{$p$-value}	\\
		\hline
		Water (L)  &       45.32 (34.70) 	&   48.29 (26.13)    &  0.519 \\
		Energy (kWh)&    2.28  (1.82)	&   2.47    (1.37)   & 0.445 \\
		Temperature (°C)&    37.09    (2.16)	&   37.67    (2.32)   & 0.089 \\
		Duration (s)&   518.77  (264.81)	&   527.20    (206.06)   & 0.812 \\
		Flow rate (L/min) &    4.41  (1.43)	&   4.83    (1.32)   & 0.039 \\
        Break time&   35.13   (39.40)	&    27.69   (32.25)   &  0.169 \\
		\hline
		Number of tenants&    1.56  (0.90)	&   1.70    (0.82)   & 0.289 \\
		Area (m$^2$)   &     46.53     (16.27)	&   48.93    (14.83)   & 0.305 \\
		Floor       &    3.55     (1.75)	&   2.55    (1.19)   & 0.000 \\
		Water meter installed &    0.79  (0.41)	&    0.89   (0.31)   & 0.066 \\
		\hline
		Households		&        86        &  92               & 178  \\
		\hline\hline
	\end{tabular}\caption{Consumption and behavior per shower in the baseline phase, along with household characteristics. Standard deviations in parentheses. The last column shows $p$-values from two-sided \textit{t}-tests.}\label{tab:shower_baseline}
\end{table}

\begin{figure}[h]
	\centering
\includegraphics[width=0.8\linewidth]{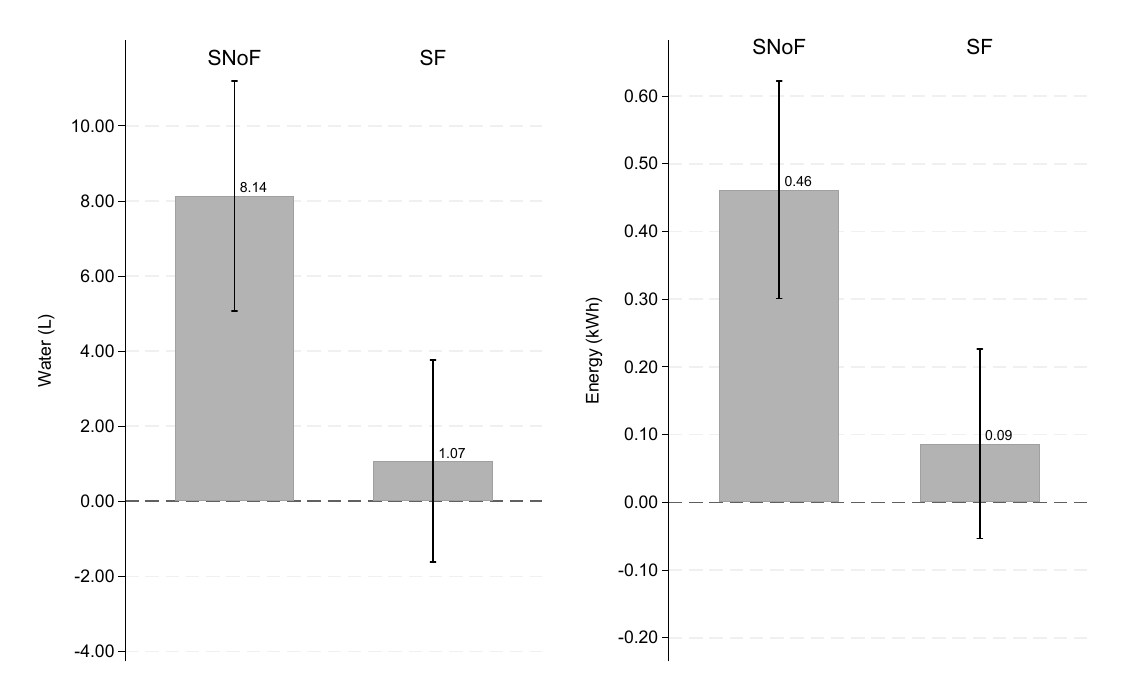}
	\caption{Mean within-household difference in water (left) and energy (right) consumption per shower by treatment. Vertical solid lines represent 95\% confidence intervals.}\label{fig:diff_showers_pre_reg}
\end{figure}

\paragraph{Effect of feedback.} Figure \ref{fig:diff_showers_pre_reg} shows the mean change in water consumption (left panel) and energy consumption (right panel) for both shower treatments. The bars represent the average within-household change in consumption between the feedback and the baseline phase. In apartments where shower heads displayed no information, we observe a significant increase in liters per shower ($+8.14$). In contrast, we do not detect a difference in shower water consumption for households that receive real-time feedback. In this group, the within-household mean difference is 1.07 liters and is not statistically different from zero at the 95\% confidence level. As expected, the pattern holds for energy consumption as well.\footnote{The amount of energy required to heat the water is calculated through the following equation: $E = (\rho c_p V (T_{\text{out}} - T_{\text{in}}))/(3600 \eta)$, where $\rho = 1000$, $c_p=4.2$, $T_{\text{in}}=10$, and $\eta=0.65$. These parameters were provided by Y-S\"{a}\"{a}ti\"{o}'s real estate unit.}

We validate the statistical significance of the above results by estimating the following equation for both water and energy consumption separately:

\begin{equation}\label{eq:feedback_effect}
r_{it} = \alpha_{i} + \beta_{1} Feedback_{it} + s_{t} + \epsilon_{it}
\end{equation}

where $r_{it}$ represents the resource (water, energy) consumption of household $i$ in shower $t$, and $Feedback_{it}$ is equal to 0 for all showers $\leq 10$, and then equal to 1 only if household $i$ is in the $SF$ treatment. We also include a household fixed effect ($\alpha_{i}$) to control for unobserved differences, and a shower fixed effect ($s_{t}$) to account for time trends in the best possible way. The coefficient $\beta_{1}$ can therefore be interpreted as the difference between the $SF$ and the $SNoF$ consumption per shower. Table \ref{tab:shower_consumption_pre_reg} reports the resulting estimates. Receiving feedback on showering behavior reduces water (energy) consumption per shower by 7 liters (0.375 kWh), resulting in a more than 13\% decrease in consumption relative to the \textit{SNoF} mean. In the last two columns, we estimate the same models separately for households with one or more tenants. We cannot reject the hypothesis that the treatment effect is of similar magnitude in both single- and multi-tenant households (\textit{F}-test: $p=0.550$ for water and $p=0.472$ for energy, respectively). Our data confirm results from other studies on the impact of real-time feedback on showering behavior \citep[e.g.,][]{TiefenbeckEtAl2018}: receiving such information reduces resource consumption, and this effect is independent of household size.

\begin{table}[h]\centering\small
\def\sym#1{\ifmmode^{#1}\else\(^{#1}\)\fi}
\begin{tabular}{l*{3}{cc}}
  \hline\hline
          &\multicolumn{1}{c}{All}&\multicolumn{1}{c}{One tenant}&\multicolumn{1}{c}{Multiple tenants}\\
  \hline
  \multicolumn{1}{l}{\textbf{Water (L)}}&&\\
  \textit{Feedback} &-7.069\sym{***} (2.05) &-7.767\sym{*} (3.14) &-5.374\sym{*} (2.48) \\
  Constant    &53.83\sym{***} (0.91) &59.74\sym{***} (1.24) &45.75\sym{***} (1.26) \\
  \hline
  \multicolumn{1}{l}{\textbf{Energy (kWh)}}&&\\
  \textit{Feedback}&-0.375\sym{***} (0.11) &-0.420\sym{*} (0.17) &-0.271\sym{*} (0.13) \\
  Constant    &2.773\sym{***} (0.05) &3.113\sym{***} (0.07) &2.306\sym{***} (0.06) \\
  \hline
  Households&      178 &      100 &       78 \\
  Observations&    12,460 &     7,000 &     5,460 \\
  \hline\hline
  \end{tabular}
  \caption{Effect of feedback. This table shows the treatment effect on water (L) and energy (kWh) consumption per shower. Standard errors clustered at the household level in parentheses. \sym{*} \(p<0.05\), \sym{**} \(p<0.01\), \sym{***} \(p<0.001\).}
  \label{tab:shower_consumption_pre_reg}
\end{table}

A central question for our empirical investigation is whether the effect of the real-time feedback attenuates over time or induces unintended behavioral responses. In principle, lower per-shower consumption could lead tenants to compensate on the extensive margin by taking fewer showers (with potential health implications) or more showers (offsetting water and energy savings). We do not find evidence supporting this. We cannot reject the hypothesis that the effect of real-time feedback is stable over time ($p=0.391$ for water and $p=0.447$ for energy), and we detect no impact on the number of showers taken ($p=0.821$).\footnote{Details on both outputs and procedures for these tests are provided in Appendix \ref{app:add_results}.} Consistent with the meter-based analysis, the addition of real-time feedback generates a further reduction in water consumption, with no detectable attenuation of the effect and no evidence of unintended consequences.

\paragraph{Behavioral channels.} We now examine through which margins the feedback intervention alters in-shower behavior. Tenants can reduce water use by shortening shower duration (i.e., turning the water off earlier), lowering the flow rate, or increasing the duration of brief shutoffs while lathering or shampooing (break time). We also study water temperature. While temperature does not directly affect volumetric consumption, it is a first-order determinant of the energy required to heat water and thus of the intervention's energy implications.\\
Table \ref{tab:shower_behavior} presents estimates from specifications analogous to equation \ref{eq:feedback_effect}, where the dependent variable is replaced with measures of in-shower behavior. The results indicate that the real-time feedback affects water use along two margins. First, treated tenants reduce shower flow rates, thereby decreasing water use. This finding is informative about concerns that low-flow fixtures constrain choice. Although the installed shower heads mechanically restrict flow, tenants who receive feedback tend to reduce the flow rate even more, suggesting that the restriction is not binding in our setting. Second, and more modestly, treated tenants shorten shower duration, with point estimates implying a reduction of roughly 30 seconds.

\begin{table}[h]\centering\small
\def\sym#1{\ifmmode^{#1}\else\(^{#1}\)\fi}
\begin{tabular}{l*{3}{cc}}
\hline\hline
          &\multicolumn{1}{c}{All}&\multicolumn{1}{c}{One tenant}&\multicolumn{1}{c}{Multiple tenants}\\
\hline
\multicolumn{1}{l}{\textbf{Temperature (°C)}}         &&&\\
\textit{Feedback} &0.100 (0.19)         &-0.159 (0.22)         &0.431 (0.31)         \\
Constant    &37.89\sym{***} (0.08)         &38.38\sym{***} (0.09)         &37.21\sym{***} (0.16)         \\
\hline
\multicolumn{1}{l}{\textbf{Duration (s)}}        &&&\\
\textit{Feedback}&-29.61\sym{*} (14.63)         &-31.10 (22.05)         &-18.58 (18.05)         \\
Constant    &570.4\sym{***} (6.48)         &619.2\sym{***} (8.69)         &502.9\sym{***} (9.13)         \\
\hline
\multicolumn{1}{l}{\textbf{Flow rate (L/min)}}         &&&\\
\textit{Feedback}&-0.387\sym{***} (0.09)         &-0.384\sym{**} (0.13)         &-0.384\sym{**} (0.14)         \\
Constant   &4.975\sym{***} (0.04)         &5.118\sym{***} (0.05)         &4.787\sym{***} (0.07)         \\
\hline
\multicolumn{1}{l}{\textbf{Break time (s)}}&&&\\
\textit{Feedback} &-1.250 (2.71)         &-2.310 (3.49)         &0.729 (4.39)         \\
Constant    &30.61\sym{***} (1.20)         &26.82\sym{***} (1.38)         &35.01\sym{***} (2.22)         \\
\hline
Households&      178         &      100         &       78         \\
Observations&    12,460         &     7,000         &     5,460         \\
\hline\hline
  \end{tabular}
  \caption{Margins of adjustment. This table shows the treatment effect on water temperature, shower duration, flow rate, and break time. Standard errors clustered at the household level in parentheses. \sym{*} \(p<0.05\), \sym{**} \(p<0.01\), \sym{***} \(p<0.001\).}
  \label{tab:shower_behavior}
\end{table}

Table \ref{tab:shower_behavior} indicates that the treatment effect is primarily driven by tenants reducing flow rate and, to a minor extent, shower time. A potential concern is that households in \textit{SF} have a higher baseline water pressure because they are, on average, one floor lower, which gives them a larger mechanical margin to decrease flow, even without real-time feedback. To address this concern, we re-estimate equation \ref{eq:feedback_effect} with flow rate as the dependent variable, restricting the sample to properties in which randomization occurred at the floor level. This within-floor comparison mitigates pre-existing differences in water pressure across treatment arms and therefore strengthens the interpretation of the estimated effect as a behavioral response to real-time feedback rather than differential adjustment capacity. The treatment effect remains statistically significant and of comparable magnitude: real-time feedback reduces flow rate by 0.447 ($p=0.002$) on average. The effect is present for both one-tenant (-0.374, $p=0.033$) and multiple-tenant households (-0.649, $p=0.027$).

\paragraph{From shower-level to household-level savings.} A useful calibration relates the treatment effect estimated at the shower level to the corresponding effect on total household water demand. Compared to the water-saving shower head without feedback, real-time feedback reduces water use by 7.07 liters per shower. Multiplying this estimate by the average number of daily showers observed in the sample (1.40) yields an implied incremental household saving of 9.90 liters per day. The corresponding estimate from the household-meter regressions in Table \ref{tab:meter_read_all_pre_reg} is 8.04 liters per day, obtained as the difference between the coefficients on \textit{SF} and \textit{SNoF}. Thus, the household-level estimates capture roughly 81\% of the water savings implied by the shower-level response. Because real-time feedback affects consumption only when the treated shower head is used, whereas the meter outcome aggregates all household water uses, the household-level effect need not map one-for-one onto the shower-level effect. The household-meter estimates should be interpreted as aggregate effects on daily household water demand and are therefore mechanically attenuated relative to the underlying per-shower behavioral response. This attenuation is reinforced by the structure of the treatment: in the \textit{SF} arm, real-time feedback becomes active only after the first ten showers, whereas the post-treatment period in the household-meter regressions begins at installation. As a result, the estimated \textit{SF}--\textit{SNoF} difference in the meter data averages over an initial interval during which the two treatments are effectively identical. For this reason, the implied pass-through from shower-level savings to household demand should be interpreted as a conservative benchmark rather than as a structural estimate. In addition, the shower and household-meter samples do not perfectly align, and the calculation of daily shower frequency relies on showers with available timestamps. Even so, the proximity of the two magnitudes is informative. It suggests that most of the water savings induced by real-time feedback at the point of use translate into lower total household demand, rather than being largely offset elsewhere in the home.\\
We further assess this interpretation by repeating the exercise on the subset of households observed in both the shower and meter data.\footnote{Please see Appendix \ref{app:add_results} for the complete regression analysis estimates.} In this matched sample, the household-meter estimates imply an incremental reduction of 7.23 liters per day, while the implied household saving from shower-level estimates amounts to 9.38 liters per day. Therefore, the estimates at the household level account for approximately 77\% of the water savings implied by the shower-level analysis. This result supports the same conclusion: the savings generated by real-time feedback at the shower level are consistent with limited offsetting when considering household water demand.\\

\section{Economic and policy implications}\label{sec:policy}

This section evaluates the economic relevance of the interventions by translating the estimated treatment effects on household water demand into private annual resource and bill savings.\footnote{This section should be interpreted as a household payback analysis rather than an assessment of the intervention's welfare effects, as in \cite{Gerster2026}.} The preceding analysis shows that the shower-head interventions reduce daily household water demand, but their policy significance depends on two additional considerations: the annual magnitude of the resulting resource savings and whether those savings are large enough to justify the acquisition cost of the devices. Table \ref{tab:annual_savings} therefore reports the implied annual savings in total, cold, and hot water use, and converts the reduction in hot-water demand into implied energy savings.\footnote{We calculate implied energy using the following equation: $E = (\rho c_p V (T_{\text{out}} - T_{\text{in}}))/(3600 \eta)$, where $\rho = 1000$, $c_p=4.2$, $T_{\text{out}}=58$, $T_{\text{in}}=10$, and $\eta=0.65$. Parameters were provided by Y-S\"{a}\"{a}ti\"{o}'s real estate unit.} Presenting the estimates in this form serves two purposes. First, it provides a natural scale for assessing the quantitative importance of the interventions. Second, it links water conservation to the energy required to supply hot water. Because the energy figures are inferred from reductions in hot-water use rather than directly metered, they should be interpreted as implied rather than directly observed energy savings.

\begin{table}[h]
\centering\small
\begin{tabular}{@{}lcccc@{}}
\toprule\toprule
\multicolumn{1}{c}{} & \textit{T} & \textit{SNoF} & \textit{SF} & \textit{SF}-\textit{SNoF} \\ \midrule
Total (m$^3$)  & -0.029 & 3.433 & 6.366 & 2.933 \\
Cold (m$^3$)   & 0.695  & 1.863 & 3.190 & 1.327 \\
Hot (m$^3$)    & -0.724 & 1.570 & 3.174 & 1.604 \\
Energy (kWh)   & -61.430 & 133.212 & 269.309 & 136.097 \\
\bottomrule\bottomrule
\end{tabular}
\caption{Annual water and implied energy savings of each intervention relative to the control arm (\textit{C}). The last column reports the difference in treatment effects between the \textit{SF} and \textit{SNoF} arms.}
\label{tab:annual_savings}
\end{table}

Table \ref{tab:annual_savings} shows that the shower timer treatment generates essentially no resource savings. Relative to the control group, it slightly increases hot-water use, implying an annual energy effect of about $-61$ kWh. By contrast, both shower-head treatments generate economically meaningful reductions in household resource use. The water-saving shower head without feedback (\textit{SNoF}) reduces total water consumption by 3.43 m$^3$ per household-year, including 1.57 m$^3$ of hot water, corresponding to an implied annual energy savings of about 133 kWh. Pairing the same device with real-time feedback (\textit{SF}) increases these savings to 6.37 m$^3$ of total water, representing a 9.9\% reduction compared to the control treatment's implied annual water consumption. Moreover, hot-water savings amount to 3.17 m$^3$, corresponding to approximately 270 kWh per year. The incremental contribution of feedback relative to \textit{SNoF} is also non-trivial: about 2.93 m$^3$ of additional total water savings and approximately 136 kWh of additional implied energy savings per household-year. 
Thus, real-time feedback yields roughly twice the annual resource savings generated by the shower head alone. For comparison, the additional energy saved by feedback from a single shower head can light an average household in the European Union for more than six months \citep[estimates based on][]{lapillonne2015}. %An average EU household uses about 239 kWh of electricity per year for lighting  These magnitudes indicate that the real-time feedback affects not only household water demand, but also the energy-relevant component of consumption through reduced demand for hot water.\\

A key policy question is whether these annual savings are large enough to justify the acquisition cost of the devices over their service life. To address this question, we exploit the fact that hot and cold water are billed separately and construct a daily expenditure measure by valuing household hot- and cold-water use at the tariffs applied in each property.\footnote{Tariffs do not vary during the study period and thus remain constant at the property level.} We then re-estimate our baseline specification using this monetary outcome. The resulting coefficients therefore measure treatment effects directly in euros per household-day (see appendix \ref{app:add_results} for the estimation output). Table \ref{tab:cost_effectiveness} reports the implied annual bill savings together with standard cost-effectiveness metrics.\footnote{The shower head has a supplier's guarantee period of five years. Therefore, the true service life is likely longer. For comparability, we apply the same lifespan horizon to the shower timer. For the Net Present Value (NPV) calculation, we assume a discount rate of 3\%.} This exercise adopts a billing perspective: it asks whether the reduction in annual water expenditures is sufficient to recover the retail cost of the intervention.

\begin{table}[h]
\centering\small
\begin{tabular}{@{}lcccc@{}}
\toprule\toprule
\multicolumn{1}{c}{} & \textit{T} & \textit{SNoF} & \textit{SF} & \textit{SF}-\textit{SNoF} \\ \midrule
Bill savings (€) & -4.84 & 22.38 & 42.29 & 19.91 \\
Upfront cost (€) & 1.5 & 170 & 170 & - \\
Payback (years)  & NA & 7.60 & 4.02 & - \\
\midrule
5-year NPV (€)   & -23.67 & -67.51 & 23.68 & 91.19 \\
7-year NPV (€)   & -31.65 & -30.57 & 93.48 & 124.05 \\
10-year NPV (€)   & -42.79 & 20.91 & 190.74 & 211.65 \\
\bottomrule\bottomrule
\end{tabular}
\caption{Annual bill savings and cost-effectiveness of each intervention relative to the control arm (\textit{C}). The last column reports the difference in treatment effects between the \textit{SF} and \textit{SNoF} arms.}
\label{tab:cost_effectiveness}
\end{table}

The monetary results closely mirror the pattern in physical savings. The timer treatment yields no meaningful monetary benefit and, if anything, slightly increases annual water expenditure. By contrast, the shower head alone reduces annual water bills by about €22.4 per household, while the shower paired with real-time feedback raises annual bill savings to €42.3. The incremental contribution of feedback is therefore €19.9 per household-year. Although these annual bill savings are modest in absolute terms, they are economically meaningful given that the intervention targets only one household activity. More importantly, they provide the relevant benchmark for evaluating whether the devices recover their acquisition cost over time.

The cost-effectiveness ranking of the interventions is clear. The shower timer is inexpensive but ineffective, and therefore does not recover even its negligible upfront cost. The water-saving shower head by itself offers significant resource savings. However, these savings are not enough to cover the retail cost over the guaranteed five-year lifespan, though they could be sufficient over its actual lifespan.\footnote{It is worth noting that the cost-effectiveness calculation for \textit{SNoF} is based on the purchase price of the smart shower head used in the study. A standard water-saving shower head that restricts flow rate but does not incorporate smart features is substantially cheaper. As a result, the \textit{SNoF} estimates likely understate the cost-effectiveness of a purely mechanical retrofit, which could deliver positive net savings over a considerably shorter payback horizon.
} By contrast, the shower head paired with real-time feedback yields positive net present value and a payback period of about four years, indicating that it is privately cost-effective on bill savings alone. The NPV estimates in the last column of Table \ref{tab:cost_effectiveness} can be interpreted as the present value of the incremental savings generated by feedback. This value is approximately €91 over five years, which is the break-even additional amount one would be willing to pay for the feedback component relative to the shower head alone. %Taken together, these results suggest that, among the interventions considered here, combining a water-saving device with real-time feedback offers the most attractive balance between effectiveness and economic viability. 
Importantly, these calculations adopt a private billing perspective and therefore capture only the reduction in water expenditures faced by households. To the extent that lower hot-water demand also reduces the real cost of supplying heated water and any associated environmental externalities, these estimates may understate the broader social value of the interventions. However, we do not quantify those additional benefits here. The results should therefore be interpreted as a conservative assessment of the interventions' economic and environmental viability.

From a policy perspective, the results imply that intervention design matters at least as much as intervention cost. A very inexpensive reminder device may appear attractive ex ante, but it is not cost-effective if it does not change behavior. A purely technological retrofit generates sizable resource savings, yet may fail to be privately attractive if its retail price is high relative to the induced reduction in water expenditure. The combination of engineering and real-time feedback performs better on both margins: it yields the largest reductions in water and implied energy use, and it is the only intervention that appears to recover its cost within the guaranteed service life. In settings where policymakers or housing providers care about both conservation and financial viability, these results favor interventions that combine a technological reduction in flow with feedback delivered at the moment of use.

%%%%%%%%%%%%%%%%%%%%%%%%%%%%%%%%%%%%%%%%%%%%%%%%%%%%%

%%%%%%%%%%%%%%%%%%%%%%%%%%%%%%%%%%%%%%%%%%%%%%%%

\section{Conclusions}\label{sec:conclusions}
This paper studies whether interventions targeted at one end use generate net reductions in total household demand. Conservation at a specific end use, indeed, might not survive if households adjust elsewhere. Using a field experiment targeting shower behavior, we compare a low-cost reminder (a shower timer), a water-saving shower head, and the same device paired with real-time feedback. We find that the shower timer does not produce meaningful savings. The water-saving shower head reduces household water demand. Real-time feedback further reduces per-shower water and energy use relative to the same device without feedback, mainly through lower flow and somewhat shorter showers. Moreover, most of those point-of-use savings appear to pass through to total household consumption rather than being fully offset elsewhere in the home.\\
The findings yield two broader implications. First, in contexts where conservation cannot rely primarily on price-based instruments, effective conservation policy may hinge on the interaction between technology and salience. Low-cost reminder-based interventions may have limited impacts, whereas technological retrofits that mechanically reduce resource use can generate meaningful savings. Pairing such retrofits with salient, point-of-use information can further amplify behavioral adjustment and improve private cost-effectiveness. Second, the results suggest that evaluations focused solely on the targeted end use may be incomplete, as savings at the point of intervention can be partly offset elsewhere in the home. For welfare and policy purposes, however, the relevant outcome is total household consumption, and our evidence indicates that end-use interventions can translate into lower aggregate demand.\\
In settings where interventions must be both operationally scalable and financially viable, approaches that combine technological constraints with salient, real-time information at the point of use appear more promising than low-cost reminder-only strategies.

\newpage
%%%%%%%%%%%%%%%%%%%%%%%%%%%%%%%%%%%%%%%%%%%%%%%%%%%%%%%%
\bibliography{biblio}

\newpage
\appendix
\addcontentsline{toc}{section}{Appendix} % Add the appendix text to the document TOC
\begin{center}
{\LARGE \textbf{Appendix for ``Beyond The Margin''}}\\
{\textbf{Andrea Albertazzi,~~Elisabetta Leni,~~Ennio Bilancini}}
\end{center}
% Start the appendix part
%\parttoc % Insert the appendix TOC

\pagenumbering{arabic}
\setcounter{footnote}{0}
\
\setcounter{table}{0}
\renewcommand{\thetable}{\thesection.\arabic{table}}
\counterwithin{table}{section}
\renewcommand\thefigure{\thesection.\arabic{figure}}
\counterwithin{figure}{section}

%---------------------------------
\section{Other Details}\label{app:other_details}
\paragraph{Properties characteristics.} Property size ranges from 15 to 119 apartments, and construction dates span the 1950s to 2022. The properties are located either in the Helsinki metropolitan area or in the city of Jyv\"{a}skyl\"{a}. Apartment sizes range from 25 to 89.5 m$^2$. Approximately 83\% include a balcony, 95\% are located in buildings with an elevator, and 99\% have no associated garden. None of the units has a private sauna.\vspace{0.5cm}

\paragraph{Delivery dates.} Start dates of the intervention for each property.
\begin{table}[H]
\centering
\begin{tabular}{@{}cc@{}}
\toprule
Treatment (number of properties) & Date \\ \midrule
\textit{T (1), SNoF (1), SF(2)} & September 25, 2023 \\
\textit{C (1), SNoF (2), SF(1)} & September 26, 2023 \\
\textit{C (2), T (1)} & September 27, 2023 \\
\textit{SF (2) SNoF (1)} & October 2, 2023 \\
\textit{SNoF (1), SF (1)} & October 3, 2023 \\
\textit{C (1), T (2),} & October 4, 2023 \\ \bottomrule
\end{tabular}
\caption{Dates of the letter and the device delivery.}
\label{tab:delivery_dates}
\end{table}

\paragraph{Letters delivered.} Letters delivered to households at the start of the intervention, for each treatment arm.

\begin{figure}[H]
	\centering
\includegraphics[width=0.9\linewidth]{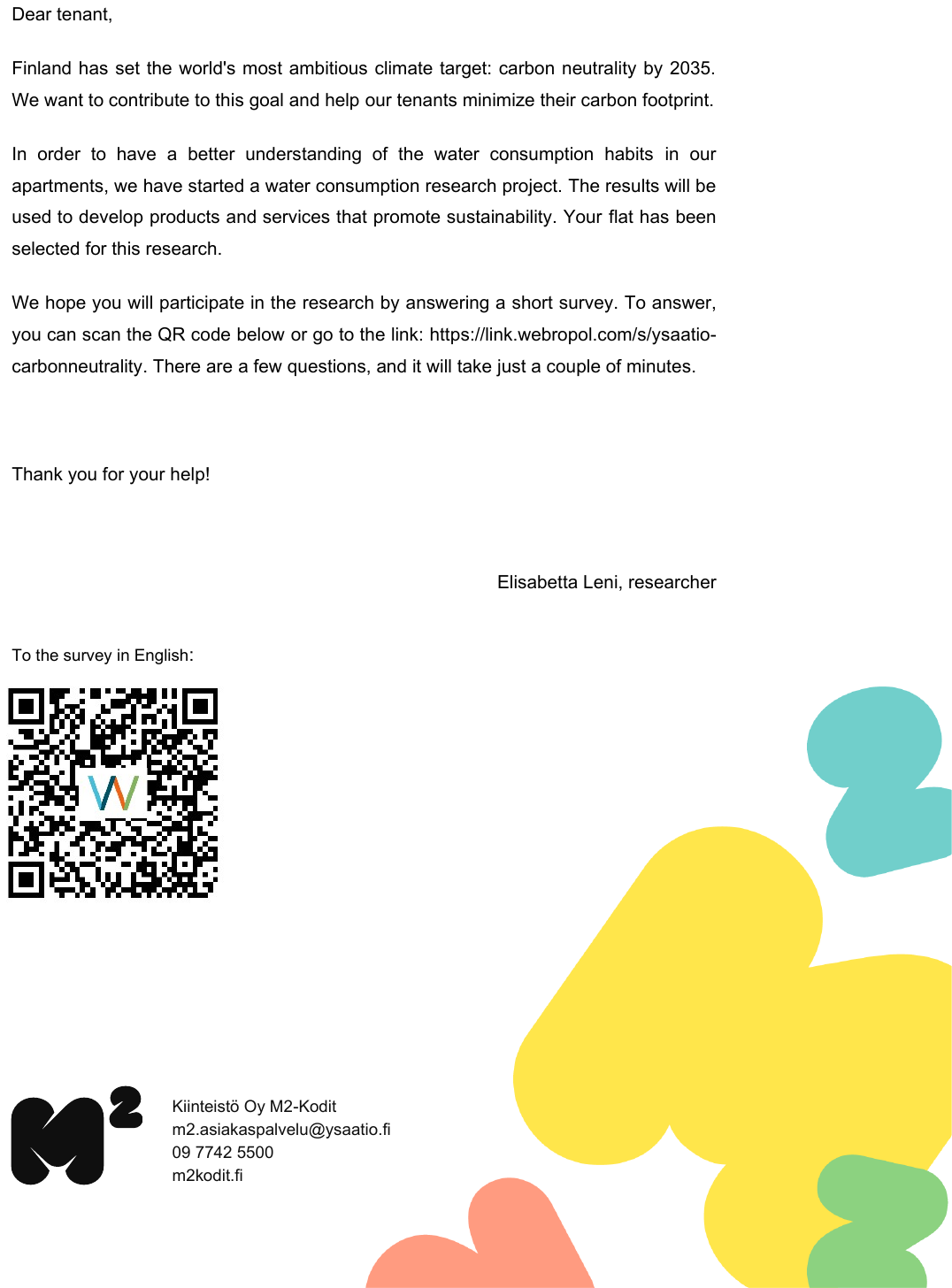}
	\caption{English version of the letter delivered to households in \textit{C}.}
\end{figure}

\begin{figure}[H]
	\centering
\includegraphics[width=0.9\linewidth]{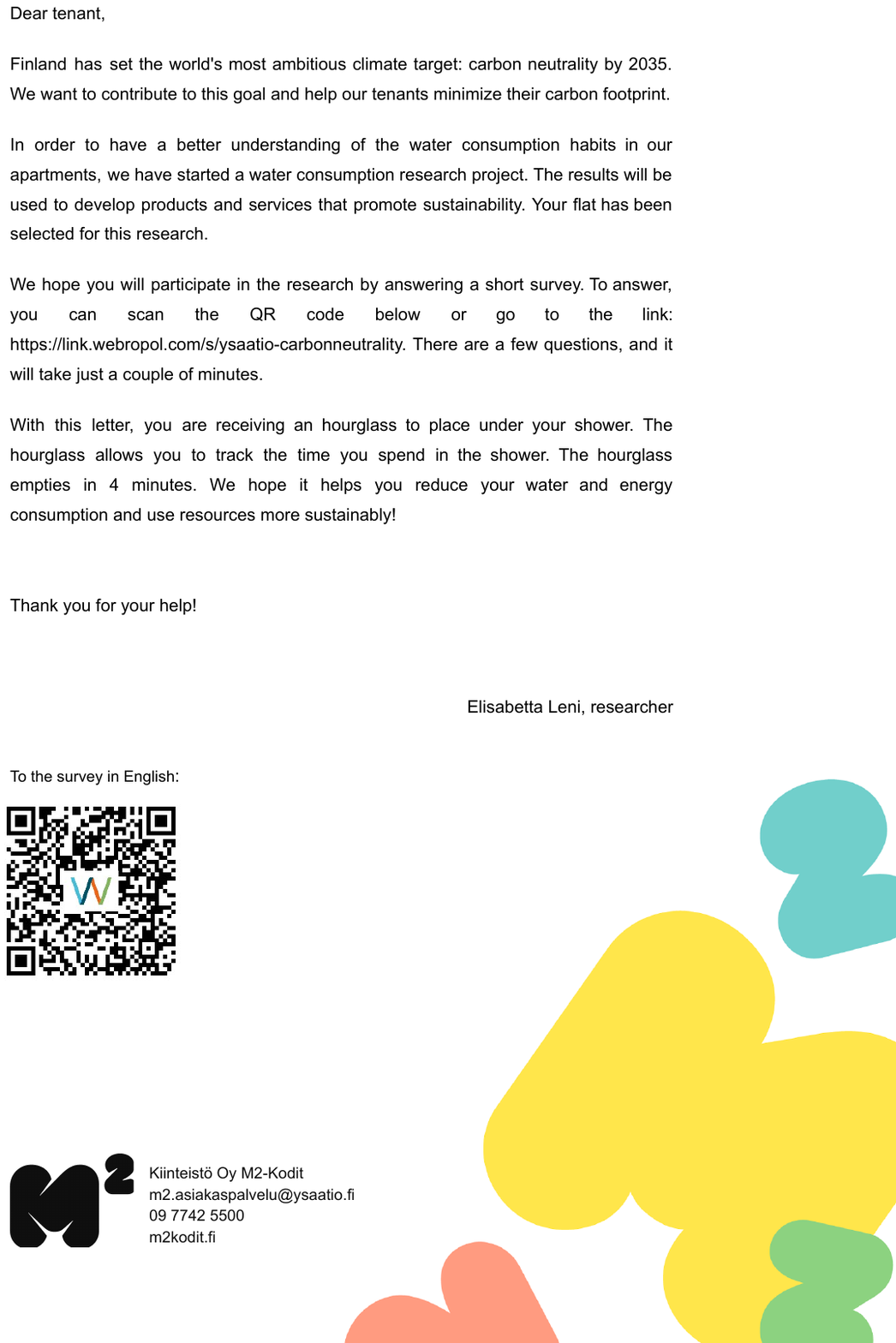}
	\caption{English version of the letter delivered to households in \textit{T}.}
\end{figure}

\begin{figure}[H]
	\centering
\includegraphics[width=0.9\linewidth]{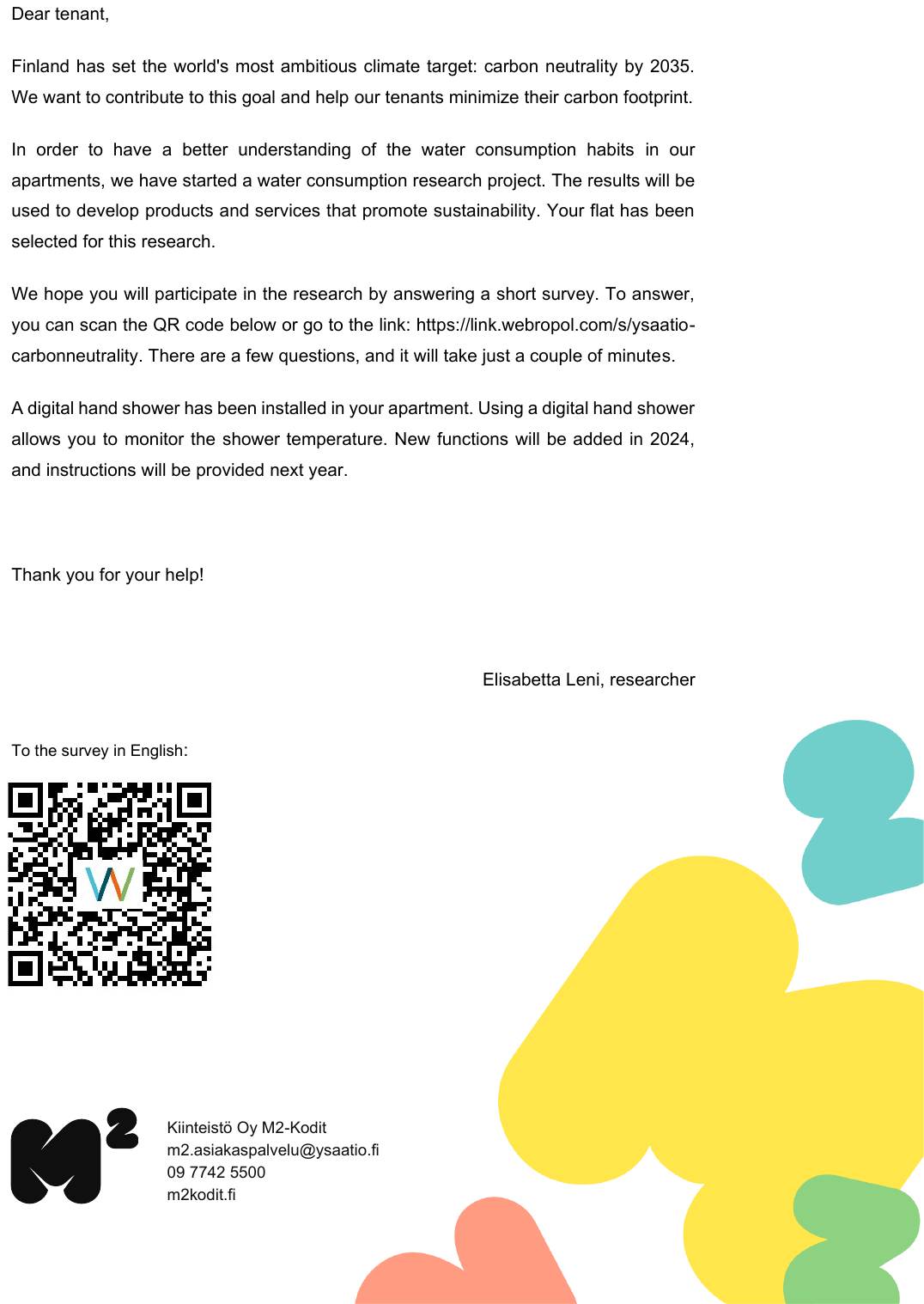}
	\caption{English version of the letter delivered to households in \textit{SNoF}.}
\end{figure}

\begin{figure}[H]
	\centering
\includegraphics[width=0.9\linewidth]{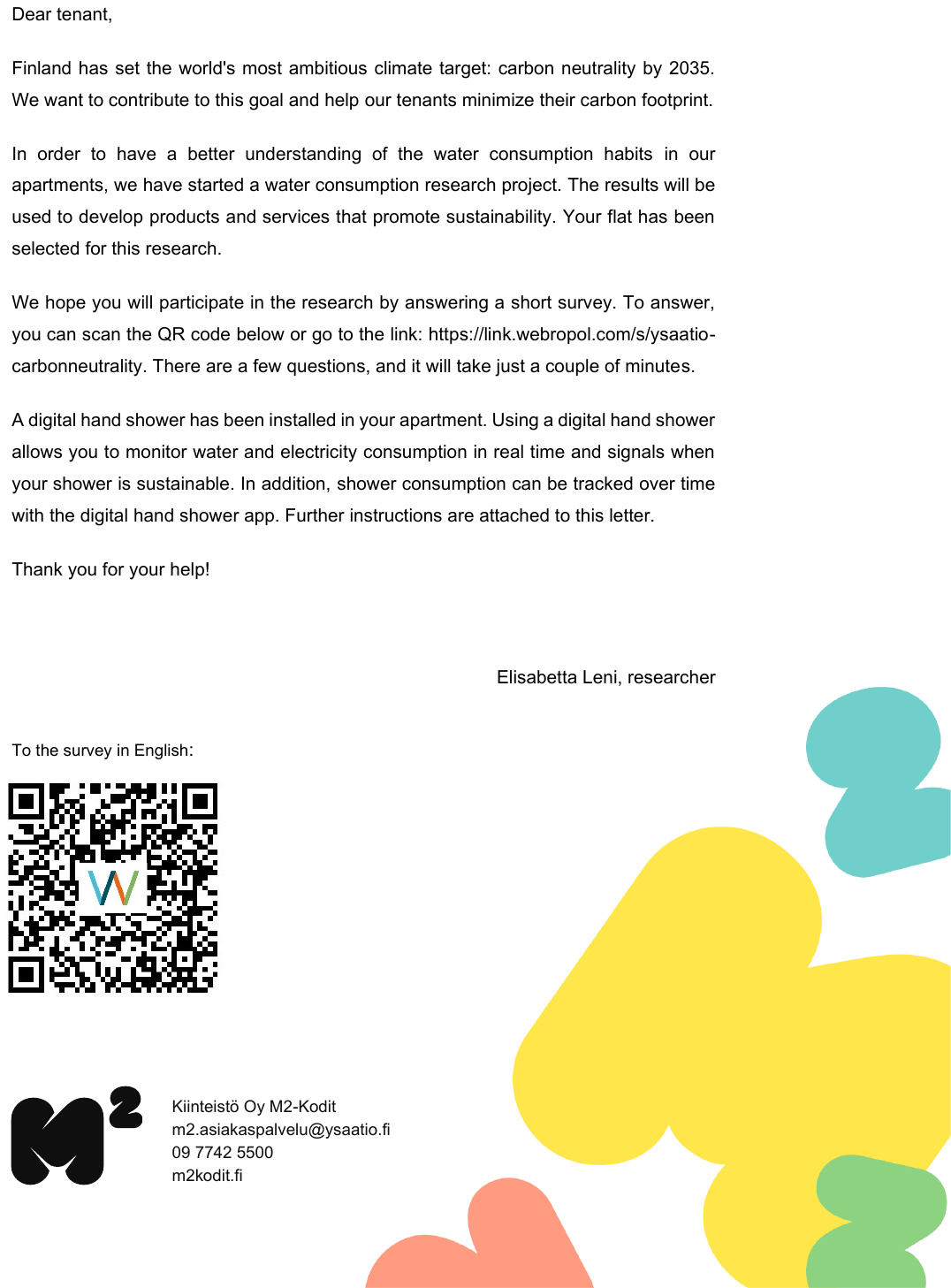}
	\caption{English version of the letter delivered to households in \textit{SF}.}
\end{figure}

\newpage
\section{Additional Results} \label{app:add_results}
In this section, we provide additional analysis that integrates our main findings.\\

\subsection{Household-level Consumption}

\begin{figure}[H]
	\centering
    \includegraphics[width=1\linewidth]{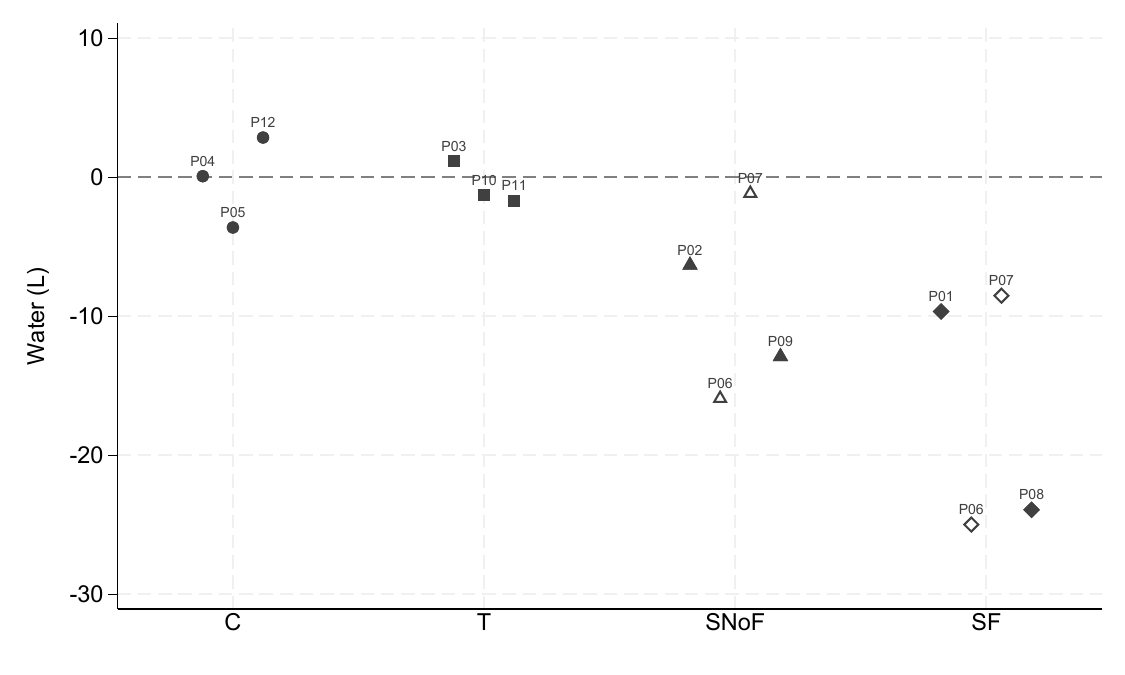}
	\caption{Mean within-property change in daily water consumption by treatment. Hollow markers denote treatment cells in the two mixed properties.}
    \label{fig:property_water_change}
\end{figure}
\vspace{0.5cm}

\paragraph{Equality of pre-trends.} In this paragraph, we test for equality of trends before the onset of interventions. We test this by estimating the following model:

\[w_{it} = \alpha_{i} + \beta_{1} Read_{t} + \beta_{2} T_{i} \times Read_{t} + \beta_{3} SNoF_{i} \times Read_{t} + \beta_{4} SF_{i} \times Read_{t} + \epsilon_{it} \]

where $w_{it}$ is the water consumption of household $i$ in meter read $t$, $Read$ represents the meter read number, and $T$, $SNoF$, and $SF$ are dummies equal to one if the household belongs to the $T$, $SNoF$, or $SF$ treatment, respectively. Hence, we test the joint hypothesis: $\beta_{2} = \beta_{3} = \beta_{4} = 0$. We also include a household fixed effect ($\alpha_{i}$). Estimates are reported in Table \ref{tab:meter_read_pre_trend_pre_reg}. From the joint F-test, we do not find evidence that pre-trends differ linearly across groups ($p = 0.525$).

\begin{table}[H]\centering\small
\def\sym#1{\ifmmode^{#1}\else\(^{#1}\)\fi}
\begin{tabular}{l*{1}{cc}}
\hline\hline
          &\multicolumn{1}{c}{Water consumption}\\
\hline
\textit{Read}  &-0.116 (0.21)         \\
\textit{T} $\times$ \textit{Read} &0.133 (0.38)         \\
\textit{SNoF} $\times$ \textit{Read} &0.303 (0.38)         \\
\textit{SF} $\times$ \textit{Read} &0.609 (0.43)         \\
Constant    &176.3\sym{***} (1.57)         \\
\hline
$\beta_{2} = \beta_{3} = \beta_{4} = 0$   &    0.524         \\
\hline
Households&       533           \\
Observations&    10,660         \\
\hline\hline
\end{tabular}
\caption{Consumption trends in the baseline (pre-intervention) phase. The regression also controls for household fixed effects. Standard errors clustered at the household level in parentheses. \sym{*} \(p<0.05\), \sym{**} \(p<0.01\), \sym{***} \(p<0.001\).}
  \label{tab:meter_read_pre_trend_pre_reg}
\end{table}
\vspace{0.5cm}

\paragraph{Matched samples.} In this paragraph, we report estimated treatment effects on daily household water consumption as in the main analysis, but restricting the shower-treatment groups to households observed in both the meter and shower samples.

\begin{table}[H]\centering\small
\def\sym#1{\ifmmode^{#1}\else\(^{#1}\)\fi}
\begin{tabular}{l*{3}{cc}}
\hline\hline
          &\multicolumn{1}{c}{Total}&\multicolumn{1}{c}{Cold}&\multicolumn{1}{c}{Hot}\\
\hline
\textit{T} &0.0806 (2.98)         &-1.903 (1.84)         &1.984 (1.51)         \\
\textit{SNoF}&-12.90\sym{*} (5.21)         &-7.591\sym{*} (3.10)         &-5.314\sym{*} (2.46)         \\
\textit{SF} &-20.14\sym{***} (4.78)         &-9.904\sym{***} (2.72)         &-10.23\sym{***} (2.40)         \\
Constant    &176.7\sym{***} (1.22)         &111.2\sym{***} (0.74)         &65.51\sym{***} (0.59)         \\
\hline
$\hat\beta_{1} = \hat\beta_{2}$ &   0.0179   &   0.079         &  0.006         \\
$\hat\beta_{1} = \hat\beta_{3}$& 0.000         &  0.005         &0.000         \\
$\hat\beta_{2} = \hat\beta_{3}$&    0.275         &    0.542         &    0.129         \\
$\hat\beta_{3} < \hat\beta_{2}$&   0.137    &         0.271    &  0.065  \\
\hline
Households &      424         &      424         &      424         \\
Observations&    33,920         &    33,920         &    33,920         \\
\hline\hline
\end{tabular}
  \caption{Effect of interventions on households' daily water consumption using matched households. Households in \textit{SNoF} and \textit{SF} are restricted to those present in both the meter and shower samples. The first column reports estimated coefficients for total daily water consumption. The last two columns report results from the estimated model for daily cold and hot water consumption only. Standard errors clustered at the household level in parentheses. \sym{*} \(p<0.05\), \sym{**} \(p<0.01\), \sym{***} \(p<0.001\).}
  \label{tab:meter_read_all_matched_pre_reg}
\end{table}
\vspace{0.5cm}

\paragraph{Four-month period.} This paragraph reports estimates of the treatment effects using all available meter data for households that did not move within the pre-registered period. Hence, we do not restrict the analysis to only 20 and 60 \textit{useful} meter reads pre- and post-intervention. We estimate the following specification:
\begin{equation}\label{eq:meters2}
    w_{it} = \alpha_{i} + \beta_{1} T_{it} + \beta_{2} SNoF_{it} + \beta_{3} SF_{it} + d_{t} + \epsilon_{it},
\end{equation}
where $w_{it}$ is household $i$'s daily water consumption in day $t$. The indicators $T_{it}$, $SNoF_{it}$, and $SF_{it}$ equal zero for all days in the baseline phase and switch to 1 thereafter for households assigned to the corresponding treatment arm. The specification includes household ($\alpha_i$) and date fixed effects ($d_t$) to absorb time-invariant heterogeneity and common shocks. Estimates are reported in Table \ref{tab:meter_read_all_4_months}.

\begin{table}[H]\centering\small
	\def\sym#1{\ifmmode^{#1}\else\(^{#1}\)\fi}
	\begin{tabular}{l*{3}{cc}}
		\hline\hline
		&\multicolumn{1}{c}{Total}&\multicolumn{1}{c}{Cold}&\multicolumn{1}{c}{Hot}\\
		\hline
\textit{T} &2.994 (3.20)         &-1.000 (1.90)         &3.994\sym{*} (1.68)         \\
\textit{SNoF} &-10.31\sym{**} (3.51)         &-5.475\sym{**} (2.06)         &-4.836\sym{**} (1.73)         \\
\textit{SF} &-16.51\sym{***} (3.90)         &-8.192\sym{***} (2.11)         &-8.313\sym{***} (2.10)         \\
Constant    &176.0\sym{***} (1.26)         &110.8\sym{***} (0.74)         &65.25\sym{***} (0.65)         \\
\hline
$\hat\beta_{1} = \hat\beta_{2}$&  0.002         &   0.061         &0.000         \\
$\hat\beta_{1} = \hat\beta_{3}$&0.000         &  0.004         &0.000         \\
$\hat\beta_{2} = \hat\beta_{3}$&    0.185         &    0.290         &    0.149         \\
$\hat\beta_{3} < \hat\beta_{2}$&  0.093  &   0.145   &  0.075  \\
\hline
Households&      573         &      573         &      573         \\
Observations&    64,192         &    64,192         &    64,192         \\
\hline\hline
\end{tabular}
\caption{Effect of interventions on households' daily water consumption using all data within the pre-registered period. The regressions also control for household and date fixed effects. Standard errors clustered at the household level in parentheses. \sym{*} \(p<0.05\), \sym{**} \(p<0.01\), \sym{***} \(p<0.001\).}
\label{tab:meter_read_all_4_months}
\end{table}
\vspace{0.5cm}

\paragraph{All data.} This paragraph reports estimates of the treatment effects using all available meter data for tenants that were originally included in the study and for up to six months after the onset of the intervention. Estimates are reported in Table \ref{tab:meter_read_all_data}.

\begin{table}[H]\centering\small
	\def\sym#1{\ifmmode^{#1}\else\(^{#1}\)\fi}
	\begin{tabular}{l*{3}{cc}}
		\hline\hline
		&\multicolumn{1}{c}{Total}&\multicolumn{1}{c}{Cold}&\multicolumn{1}{c}{Hot}\\
		\hline
\textit{T} &3.204 (3.46)         &-1.832 (2.13)         &5.036\sym{**} (1.79)         \\
\textit{SNoF} &-10.57\sym{**} (3.59)         &-6.947\sym{**} (2.19)         &-3.627\sym{*} (1.79)         \\
\textit{SF} &-15.06\sym{***} (4.13)         &-7.844\sym{***} (2.26)         &-7.214\sym{**} (2.27)         \\
Constant    &177.1\sym{***} (1.59)         &110.9\sym{***} (0.94)         &66.19\sym{***} (0.83)         \\
\hline
$\hat\beta_{1} = \hat\beta_{2}$&  0.001         &   0.042        &0.000        \\
$\hat\beta_{1} = \hat\beta_{3}$&0.000         &   0.021         &0.000        \\
$\hat\beta_{2} = \hat\beta_{3}$ &0.334         &    0.732         &    0.143         \\
$\hat\beta_{3} < \hat\beta_{2}$& 0.167& 0.366 & 0.072 \\
\hline
Households&      590         &      590         &      590         \\
Observations&   117,980         &   117,980         &   117,980         \\
\hline\hline
\end{tabular}
\caption{Effect of interventions on households' daily water consumption using all data up to six months after the onset of intervention. The regressions also control for household and date fixed effects. Standard errors clustered at the household level in parentheses. \sym{*} \(p<0.05\), \sym{**} \(p<0.01\), \sym{***} \(p<0.001\).}
\label{tab:meter_read_all_data}
\end{table}
\vspace{0.5cm}

\paragraph{Bill savings.} This paragraph reports estimated treatment effects as in the main analysis, but with daily water expenditure as the dependent variable. Estimates are reported in Table \ref{tab:meter_savings}.

\begin{table}[H]\centering\small
	\def\sym#1{\ifmmode^{#1}\else\(^{#1}\)\fi}
	\begin{tabular}{l*{1}{cc}}
		\hline\hline
		&\multicolumn{1}{c}{Costs (€)}\\
		\hline
		\textit{T} &0.0133 (0.02)         \\
\textit{SNoF} &-0.0613\sym{*} (0.02)         \\
\textit{SF}&-0.116\sym{***} (0.03)         \\
Constant    &1.021\sym{***} (0.01)         \\
\hline
$\hat\beta_{1} = \hat\beta_{2}$&  0.009         \\
$\hat\beta_{1} = \hat\beta_{3}$& 0.000         \\
$\hat\beta_{2} = \hat\beta_{3}$&   0.097         \\
$\hat\beta_{3} < \hat\beta_{2}$& 0.048 \\
\hline
Apartments&      533         \\
Observations&    42,640         \\
\hline\hline
\end{tabular}
\caption{Effect of interventions on households' daily water expenditure (€). Standard errors clustered at the household level in parentheses. \sym{*} \(p<0.05\), \sym{**} \(p<0.01\), \sym{***} \(p<0.001\).}
\label{tab:meter_savings}
\end{table}
\vspace{1cm}

%%%%%%%%%%%%%%%%%%%%%%%%%%%%%%%%%%%%%%%%%%%%%%5
\subsection{Shower-level Consumption}

\paragraph{Stability of treatment effect.} To test whether the effect of the feedback attenuates over time, we follow \cite{TiefenbeckEtAl2018} and estimate the following model:

\[R_{it} = \alpha_{i} + \beta_{1} Feedback_{it} + \beta_2 Feedback_{it} \times x_{it} + s_{t} + \epsilon_{it} \]

where $R_{it}$ represents the resource (water, energy) consumption of household $i$ in shower $t$, $Feedback_{it}$ is equal to 0 for all showers $\leq 10$, and then equal to 1 only if household $i$ is in the Feedback treatment, and $x_{it}$ represents the fraction of the treatment phase completed: $x_{it} = (t - 11)/(70 - 11)$ for $t>= 11$, and $x_{it} = 0$ for $t<11$. Thus, $\beta_2$ measures variations in the treatment effect by fitting a linear trend between the first and the last shower of the treatment phase. We also include a household ($\alpha_{i}$) and shower fixed effect ($s_{t}$).\\
Estimation results are reported in Table \ref{tab:shower_trends}. For both water and energy, we do not detect any significant change in the treatment effect over the time considered, as the coefficient $\beta_2$ is not statistically significant at any conventional level ($p=0.391$ and $p=0.447$ for water and energy, respectively).

\begin{table}[H]\centering\small
\def\sym#1{\ifmmode^{#1}\else\(^{#1}\)\fi}
\begin{tabular}{l*{2}{cc}}
\hline\hline
        
          &\multicolumn{1}{c}{Water (L)}&\multicolumn{1}{c}{Energy (kWh)}\\
\hline
\textit{Feedback}&-8.299\sym{***} (2.19)         &-0.433\sym{***} (0.11)         \\
$Feedback \times x_{it}$&2.462 (2.86)         &0.116 (0.15)         \\
Constant    &53.83\sym{***} (0.91)         &2.773\sym{***} (0.05)         \\
\hline
Households &      178         &      178         \\
Observations&    12,460         &    12,460         \\
\hline\hline
\end{tabular}
\caption{Stability of treatment effect. This table shows the temporal stability of the treatment effect on water (L) and energy (kWh) consumption per shower. Standard errors clustered at the household level in parentheses. \sym{*} \(p<0.05\), \sym{**} \(p<0.01\), \sym{***} \(p<0.001\).}
  \label{tab:shower_trends}
\end{table}

\vspace{0.5cm}

\paragraph{Negative externalities.} Tenants who receive real-time feedback might change their shower frequency. We test this hypothesis by estimating the following equation:

\[TS_i = \alpha + \beta_{1} Feedback_{i} + Start_i +  \epsilon_{i} \]

where $TS_i$ represents the total number of showers taken by household $i$ during the study period, $Feedback_{i}$ is a dummy indicating the assignment to the feedback treatment, and $Start_i$ is an intervention-start-date fixed effect. The intervention starting date is the day when households were informed about the study and shower heads were installed. Because shower data are observed only from the installation date onward, differences in such a date translate into differences in exposure to the intervention, potentially affecting the total number of showers recorded. The fixed effect $Start_i$ controls for this contingency.\\
As reported in Table \ref{tab:num_of_showers}, we find no significant evidence for the treatment to induce households to take more or fewer showers.

\begin{table}[H]\centering\small
\def\sym#1{\ifmmode^{#1}\else\(^{#1}\)\fi}
\begin{tabular}{l*{3}{cc}}
\hline\hline
          &\multicolumn{1}{c}{All}&\multicolumn{1}{c}{One tenant}&\multicolumn{1}{c}{Multiple tenants}\\
\hline
\textit{Feedback}  &2.456 (10.62)         &-7.714 (12.14)         &16.43 (22.12)         \\
Constant   &158.4\sym{***} (8.22)         &142.7\sym{***} (8.22)         &176.3\sym{***} (14.43)         \\
\hline
\hline
Observations&      178         &      100         &       78         \\
  \hline\hline
  \end{tabular}
  \caption{Treatment effect on total number of showers taken (linear regression). Regressions also control for the day the shower head was installed. Standard errors clustered at the building level in parentheses. \sym{*} \(p<0.05\), \sym{**} \(p<0.01\), \sym{***} \(p<0.001\).}
  \label{tab:num_of_showers}
\end{table}

\vspace{0.5cm}

\paragraph{Financial incentives.} A policy-relevant concern is whether behavioral responses depend on households' incentives to internalize consumption costs. In our setting, some apartments have a private water meter (and thus face consumption-based pricing), whereas others pay based on a share of total property consumption. To test whether the response to the feedback varies with financial incentives, we estimate the following equation:

\[R_it = \alpha_{i} + \beta_{1} Feedback_{it} + \beta_{2}  Feedback_{it} \times Meter_{i} + s_{t} + \epsilon_{it} \]

where $Meter$ equals 1 if the household has a private water meter installed, and 0 otherwise. The coefficient $\beta_{2}$ therefore represents the additional treatment effect on households that face consumption-based pricing. Estimates are reported in Table \ref{tab:meter_installed}. Consistent with the feedback mechanism operating through salience rather than differential financial incentives, we do not detect statistically meaningful heterogeneity in treatment effects by metering status.

\begin{table}[H]\centering\small
\def\sym#1{\ifmmode^{#1}\else\(^{#1}\)\fi}
\begin{tabular}{l*{2}{cc}}
\hline\hline
          &\multicolumn{1}{c}{Water (L)}&\multicolumn{1}{c}{Energy (kWh)}\\
\hline
$Feedback$ &-8.090\sym{*} (3.16)         &-0.403\sym{*} (0.16)         \\
$Feedback \times Meter$&1.146 (3.14)         &0.031 (0.16)         \\
Constant    &53.83\sym{***} (0.91)         &2.773\sym{***} (0.05)         \\
\hline
Households & 178 & 178 \\
Observations&    12,460         &    12,460         \\
\hline\hline
\end{tabular}
\caption{Financial incentives. This table shows whether the treatment effect depends on consumption-based pricing. Standard errors clustered at the household level in parentheses. \sym{*} \(p<0.05\), \sym{**} \(p<0.01\), \sym{***} \(p<0.001\).}\label{tab:meter_installed}
\end{table}

\vspace{0.5cm}

\paragraph{Spillovers.} 

Here, we assess whether the intervention generates spillovers onto households that never receive feedback. Spillovers are most plausible in properties where randomization occurred at the floor level, because treated and control households share common spaces and may exchange information about the feedback. Focusing on households in the $SNoF$ group, we compare those in floor-randomized buildings to those in buildings assigned entirely to the $SNoF$ condition. To this purpose, we estimate the following equation only on households that belong to \textit{SNoF}, and therefore never receive real-time feedback:

\[R_{it} = \alpha_{i} + \beta_{1} After_{it} \times Mixed_{i} + s_{t} + \epsilon_{it} \]

where $After$ is equal to 1 if the observation belongs to the intervention phase, and 0 to the baseline phase. $Mixed$ is equal to 1 if the household belongs to a property where randomization occurred at the floor level, 0 otherwise. 
Table \ref{tab:spillovers} reports the estimated coefficients. We find no evidence of spillovers for either water or energy consumption.

\begin{table}[H]\centering\small
\def\sym#1{\ifmmode^{#1}\else\(^{#1}\)\fi}
\begin{tabular}{l*{2}{cc}}
\hline\hline
          &\multicolumn{1}{c}{Water (L)}&\multicolumn{1}{c}{Energy (kWh)}\\
\hline
$After \times Mixed$&-1.256 (3.38)         &-0.0371 (0.18)         \\
Constant    &52.74\sym{***} (1.18)         &2.690\sym{***} (0.06)         \\
\hline
Households & 86 & 86 \\
Observations& 6,020 & 6,020 \\
\hline\hline
\end{tabular}
\caption{Spillovers. This table shows whether the treatment effect extends to non-treated households in properties where randomization occurred at the floor level. Standard errors clustered at the household level in parentheses. \sym{*} \(p<0.05\), \sym{**} \(p<0.01\), \sym{***} \(p<0.001\).}\label{tab:spillovers}
\end{table}
\vspace{0.5cm}

\paragraph{Matched samples.} In this paragraph, we report the treatment effect of feedback on shower water consumption, restricted to households that are present in both the meter and shower samples.

\begin{table}[H]\centering\small
	\def\sym#1{\ifmmode^{#1}\else\(^{#1}\)\fi}
	\begin{tabular}{l*{3}{cc}}
		\hline\hline
		&\multicolumn{1}{c}{All}&\multicolumn{1}{c}{One tenant}&\multicolumn{1}{c}{Multiple tenants}\\
		\hline
		\multicolumn{1}{l}{\textbf{Water (L)}}&&\\
\textit{Feedback} &-6.791\sym{**} (2.44)         &-6.907\sym{*} (3.44)         &-6.160 (3.38)         \\
Constant    &55.15\sym{***} (1.14)         &60.47\sym{***} (1.39)         &46.54\sym{***} (1.91)         \\
\hline
\multicolumn{1}{l}{\textbf{Energy (kWh)}}&&\\
\textit{Feedback}&-0.368\sym{**} (0.13)         &-0.376\sym{*} (0.18)         &-0.323 (0.18)         \\
Constant    &2.853\sym{***} (0.06)         &3.157\sym{***} (0.07)         &2.357\sym{***} (0.10)         \\
\hline
Apartments&      136         &       83         &       53         \\
Observations&     9,520         &     5,810         &     3,710         \\
\hline\hline
\end{tabular}
\caption{Effect of feedback, using matched households only. This table shows the treatment effect on water (L) and energy (kWh) consumption per shower. Standard errors clustered at the household level in parentheses. \sym{*} \(p<0.05\), \sym{**} \(p<0.01\), \sym{***} \(p<0.001\).}
\label{tab:shower_consumption_matched}
\end{table}
\vspace{0.5cm}

\paragraph{All data.} This paragraph reports estimates of the feedback effect on shower consumption using all available shower data in the period of interest. We therefore re-estimate our main specification without the pre-registered restriction on the number of showers per household: 10 in the baseline phase, and 60 in the intervention phase. Estimates are reported in Table \ref{tab:shower_consumption_all_data}.

\begin{table}[H]\centering\small
\def\sym#1{\ifmmode^{#1}\else\(^{#1}\)\fi}
\begin{tabular}{l*{3}{cc}}
  \hline\hline
          &\multicolumn{1}{c}{All}&\multicolumn{1}{c}{One tenant}&\multicolumn{1}{c}{Multiple tenants}\\
  \hline
  \multicolumn{1}{l}{\textbf{Water (L)}}&&\\
  \textit{Feedback} &-5.931\sym{***} (1.78)         &-6.429\sym{*} (2.61)         &-5.200\sym{*} (2.27)         \\
  Constant    &50.93\sym{***} (0.90)         &56.60\sym{***} (1.11)         &45.45\sym{***} (1.32)         \\
  \hline
  \multicolumn{1}{l}{\textbf{Energy (kWh)}}&&\\
  \textit{Feedback}&-0.349\sym{***} (0.10)         &-0.387\sym{**} (0.15)         &-0.290\sym{*} (0.12)         \\
  Constant    &2.641\sym{***} (0.05)         &2.957\sym{***} (0.06)         &2.328\sym{***} (0.07)         \\
  \hline
  Households&      282         &      173         &      109         \\
  Observations&    36,329         &    17,425         &    18,860         \\
  \hline\hline
  \end{tabular}
  \caption{Effect of feedback using all data. This table shows the treatment effect on water (L) and energy (kWh) consumption per shower. Standard errors clustered at the household level in parentheses. \sym{*} \(p<0.05\), \sym{**} \(p<0.01\), \sym{***} \(p<0.001\).}
  \label{tab:shower_consumption_all_data}
\end{table}

\subsection{Survey}
Summary statistics of selected survey questions.

\begin{table}[H]
\centering
\resizebox{\textwidth}{!}{%
\begin{tabular}{@{}lccccccccccc@{}}
\toprule
& \multicolumn{11}{c}{Treatment} \\ 
\midrule
& \multicolumn{2}{c}{\textit{C}} 
&  & \multicolumn{2}{c}{\textit{T}} 
&  & \multicolumn{2}{c}{\textit{SNoF}} 
&  & \multicolumn{2}{c}{\textit{SF}} \\ 
\cmidrule(lr){2-3} 
\cmidrule(lr){5-6} 
\cmidrule(lr){8-9} 
\cmidrule(l){11-12} 
& \multicolumn{1}{c}{Before} 
& \multicolumn{1}{c}{After}
&  
& \multicolumn{1}{c}{Before} 
& \multicolumn{1}{c}{After} 
&  
& \multicolumn{1}{c}{Before} 
& \multicolumn{1}{c}{After} 
&  
& \multicolumn{1}{c}{Before} 
& \multicolumn{1}{c}{After} \\
\midrule

\begin{tabular}[c]{@{}l@{}}
I actively monitor the energy and \\
water consumption of my home.\\
(1 = completely disagree, 10 = completely agree)
\end{tabular}
& \multicolumn{2}{c}{
    \begin{tabular}[c]{@{}cc@{}}
    \begin{tabular}[c]{@{}c@{}}6.471\\ (2.37)\end{tabular}
    &
    \begin{tabular}[c]{@{}c@{}}6.824\\ (2.58)\end{tabular} \\
    \multicolumn{2}{c}{0.353} \\
    \multicolumn{2}{c}{$p=0.470$}
    \end{tabular}
}
&
& \multicolumn{2}{c}{
    \begin{tabular}[c]{@{}cc@{}}
    \begin{tabular}[c]{@{}c@{}}7\\ (2.63)\end{tabular}
    &
    \begin{tabular}[c]{@{}c@{}}7.333\\ (2.64)\end{tabular} \\
    \multicolumn{2}{c}{0.333} \\
    \multicolumn{2}{c}{$p=0.627$}
    \end{tabular}
}
&
& \multicolumn{2}{c}{
    \begin{tabular}[c]{@{}cc@{}}
    \begin{tabular}[c]{@{}c@{}}6.167\\ (2.36)\end{tabular}
    &
    \begin{tabular}[c]{@{}c@{}}5.722\\ (2.11)\end{tabular} \\
    \multicolumn{2}{c}{-0.444} \\
    \multicolumn{2}{c}{$p=0.457$}
    \end{tabular}
}
&
& \multicolumn{2}{c}{
    \begin{tabular}[c]{@{}cc@{}}
    \begin{tabular}[c]{@{}c@{}}5.65\\ (2.13)\end{tabular}
    &
    \begin{tabular}[c]{@{}c@{}}6.8\\ (2.07)\end{tabular} \\
    \multicolumn{2}{c}{1.15} \\
    \multicolumn{2}{c}{$p=0.013$}
    \end{tabular}
}
\\

\addlinespace

\begin{tabular}[c]{@{}l@{}}
In your decision about water and\\
energy consumption: how important\\
are economic considerations\\
(saving on utility bills)?\\
(1 = not important, 10 = extremely important)
\end{tabular}
& \multicolumn{2}{c}{
    \begin{tabular}[c]{@{}cc@{}}
    \begin{tabular}[c]{@{}c@{}}7.529\\ (1.81)\end{tabular}
    &
    \begin{tabular}[c]{@{}c@{}}8.118\\ (1.54)\end{tabular} \\
    \multicolumn{2}{c}{0.588} \\
    \multicolumn{2}{c}{$p=0.250$}
    \end{tabular}
}
&
& \multicolumn{2}{c}{
    \begin{tabular}[c]{@{}cc@{}}
    \begin{tabular}[c]{@{}c@{}}7.167\\ (2.29)\end{tabular}
    &
    \begin{tabular}[c]{@{}c@{}}7.833\\ (2.29)\end{tabular} \\
    \multicolumn{2}{c}{0.667} \\
    \multicolumn{2}{c}{$p=0.418$}
    \end{tabular}
}
&
& \multicolumn{2}{c}{
    \begin{tabular}[c]{@{}cc@{}}
    \begin{tabular}[c]{@{}c@{}}7.944\\ (2.15)\end{tabular}
    &
    \begin{tabular}[c]{@{}c@{}}7.556\\ (2.12)\end{tabular} \\
    \multicolumn{2}{c}{-0.389} \\
    \multicolumn{2}{c}{$p=0.453$}
    \end{tabular}
}
&
& \multicolumn{2}{c}{
    \begin{tabular}[c]{@{}cc@{}}
    \begin{tabular}[c]{@{}c@{}}7.4\\ (1.88)\end{tabular}
    &
    \begin{tabular}[c]{@{}c@{}}7.85\\ (1.63)\end{tabular} \\
    \multicolumn{2}{c}{0.45} \\
    \multicolumn{2}{c}{$p=0.243$}
    \end{tabular}
}
\\

\addlinespace

\begin{tabular}[c]{@{}l@{}}
In your decision about water and\\
energy consumption: how important\\
are environmental considerations\\
(saving natural resources)?\\
(1 = not important, 10 = extremely important)
\end{tabular}
& \multicolumn{2}{c}{
    \begin{tabular}[c]{@{}cc@{}}
    \begin{tabular}[c]{@{}c@{}}6.941\\ (2.84)\end{tabular}
    &
    \begin{tabular}[c]{@{}c@{}}8.176\\ (1.85)\end{tabular} \\
    \multicolumn{2}{c}{1.235} \\
    \multicolumn{2}{c}{$p=0.034$}
    \end{tabular}
}
&
& \multicolumn{2}{c}{
    \begin{tabular}[c]{@{}cc@{}}
    \begin{tabular}[c]{@{}c@{}}8.167\\ (1.75)\end{tabular}
    &
    \begin{tabular}[c]{@{}c@{}}8.333\\ (2.06)\end{tabular} \\
    \multicolumn{2}{c}{0.167} \\
    \multicolumn{2}{c}{$p=0.689$}
    \end{tabular}
}
&
& \multicolumn{2}{c}{
    \begin{tabular}[c]{@{}cc@{}}
    \begin{tabular}[c]{@{}c@{}}6.889\\ (1.97)\end{tabular}
    &
    \begin{tabular}[c]{@{}c@{}}7.333\\ (2.45)\end{tabular} \\
    \multicolumn{2}{c}{0.444} \\
    \multicolumn{2}{c}{$p=0.526$}
    \end{tabular}
}
&
& \multicolumn{2}{c}{
    \begin{tabular}[c]{@{}cc@{}}
    \begin{tabular}[c]{@{}c@{}}6.8\\ (2.78)\end{tabular}
    &
    \begin{tabular}[c]{@{}c@{}}7.9\\ (1.97)\end{tabular} \\
    \multicolumn{2}{c}{1.1} \\
    \multicolumn{2}{c}{$p=0.034$}
    \end{tabular}
}
\\

\addlinespace

\begin{tabular}[c]{@{}l@{}}
How much water is consumed in\\
your flat every day, on average per person?\\
Guess if you don't know.
\end{tabular}
& \multicolumn{2}{c}{
    \begin{tabular}[c]{@{}cc@{}}
    \begin{tabular}[c]{@{}c@{}}96.471\\ (56.23)\end{tabular}
    &
    \begin{tabular}[c]{@{}c@{}}71.176\\ (52.43)\end{tabular} \\
    \multicolumn{2}{c}{-25.294} \\
    \multicolumn{2}{c}{$p=0.089$}
    \end{tabular}
}
&
& \multicolumn{2}{c}{
    \begin{tabular}[c]{@{}cc@{}}
    \begin{tabular}[c]{@{}c@{}}95.833\\ (64.59)\end{tabular}
    &
    \begin{tabular}[c]{@{}c@{}}101.667\\ (68.34)\end{tabular} \\
    \multicolumn{2}{c}{5.833} \\
    \multicolumn{2}{c}{$p=0.664$}
    \end{tabular}
}
&
& \multicolumn{2}{c}{
    \begin{tabular}[c]{@{}cc@{}}
    \begin{tabular}[c]{@{}c@{}}96.111\\ (62.04)\end{tabular}
    &
    \begin{tabular}[c]{@{}c@{}}87.778\\ (46.72)\end{tabular} \\
    \multicolumn{2}{c}{-8.333} \\
    \multicolumn{2}{c}{$p=0.591$}
    \end{tabular}
}
&
& \multicolumn{2}{c}{
    \begin{tabular}[c]{@{}cc@{}}
    \begin{tabular}[c]{@{}c@{}}89.5\\ (38.45)\end{tabular}
    &
    \begin{tabular}[c]{@{}c@{}}108\\ (45.38)\end{tabular} \\
    \multicolumn{2}{c}{18.5} \\
    \multicolumn{2}{c}{$p=0.102$}
    \end{tabular}
}
\\

\midrule
Households 
& \multicolumn{2}{c}{\textit{17}}  
& 
& \multicolumn{2}{c}{\textit{12}}  
&  
& \multicolumn{2}{c}{\textit{18}} 
&   
& \multicolumn{2}{c}{\textit{20}} \\ 

\bottomrule 
\end{tabular}
}
\caption{Means and standard deviations in parentheses before and after each intervention. Within-treatment differences between after and before interventions are reported for each question along with their statistical significance (two-sided t-test).}
\label{tab:survey}
\end{table}

\end{document}